\documentclass[a4paper,fleqn,usenatbib]{mnras}

\usepackage[T1]{fontenc}
\usepackage{ae,aecompl}

\usepackage{graphicx}	
\usepackage{amsmath}	
\usepackage{amssymb}	
\graphicspath{{plots/}}
\usepackage[nolist]{acronym}
\usepackage{upgreek}
\usepackage[separate-uncertainty=true]{siunitx} 
\usepackage[flushleft]{threeparttable}
\usepackage{xcolor}

\title[Massive quiescent galaxies at $z\sim 3$]{Massive quiescent galaxies at $z\sim 3$: a comparison of selection, stellar population and structural properties with simulation predictions}

\author[P. Lustig et al.]{Peter Lustig,$^{1,2,3,4}$\thanks{E-mail: plustig@sissa.it}
Veronica Strazzullo,$^{4,5,3}$
Rhea-Silvia Remus,$^2$
Chiara D'Eugenio,$^{6,7}$
Emanuele Daddi,$^{8}$\newauthor
Andreas Burkert,$^{2}$
Gabriella De Lucia,$^{4}$
Ivan Delvecchio,$^{5}$
Klaus Dolag,$^{2}$
Fabio Fontanot, $^{3,4}$\newauthor
Raphael Gobat,$^{9,10}$
Joseph J. Mohr,$^{1,2}$
Masato Onodera,$^{11,12}$
Maurilio Pannella,$^{13,4}$\newauthor
and Annalisa Pillepich$^{14}$
\\
$^{1}$Max Planck Institute for Extraterrestrial Physics, Giessenbachstrasse, 85748 Garching, Germany\\
$^{2}$Faculty of Physics, Ludwig-Maximilians-Universität, Scheinerstr. 1, 81679 Munich, Germany\\
$^{3}$IFPU - Institute for Fundamental Physics of the Universe, Via Beirut 2, 34151 Trieste, Italy\\
$^{4}$INAF – Osservatorio Astronomico di Trieste, Via Tiepolo 11, 34131 Trieste, Italy\\
$^{5}$INAF – Osservatorio Astronomico di Brera, Via Brera 28, I-20121, Milano, Italy \& Via Bianchi 46, I-23807, Merate, Italy\\
$^{6}$Instituto de Astrofísica de Canarias (IAC), E-38205 La Laguna, Tenerife, Spain\\
$^{7}$Universidad de La Laguna, Dpto. Astrofísica, E-38206 La Laguna, Tenerife, Spain\\
$^{8}$Universit\'e Paris-Saclay, Universit\'e Paris Cit\'e, CEA, CNRS, AIM, 91191, Gif-sur-Yvette, France\\
$^{9}$Instituto de Física, Pontificia Universidad Católica de Valparaíso,
Casilla 4059, Valparaíso, Chile\\
$^{10}$INAF-Osservatorio Astronomico di Capodimonte, Via Moiariello
16, 80131 Napoli, Italy\\
$^{11}$The Graduate University for Advanced Studies, SOKENDAI, 2-21-1 Osawa, Mitaka, Tokyo, 181-8588, Japan\\
$^{12}$Subaru Telescope, National Astronomical Observatory of Japan, National Institutes of Natural Sciences, 650 North A’ohoku Place, Hilo, HI 96720, USA\\
$^{13}$Università di Trieste, Dipartimento di Fisica, Sezione di Astronomia, via Tiepolo 11,
34143 Trieste, Italy\\
$^{14}$Max-Planck-Institut für Astronomie, Königstuhl 17, 69117 Heidelberg, Germany
}

\date{Accepted XXX. Received YYY; in original form ZZZ}

\pubyear{2022}

\DeclareSIUnit\angstrom{\text {Å}}
\usepackage{newtxtext,newtxmath}
\begin{document}
\newcommand{\Ahz}{A_{\mathrm{H, obs}}}
\newcommand{\Au}{A_{\mathrm{U}}}
\newcommand{\Auv}{A_{\mathrm{UV}}}
\newcommand{\Aj}{A_{\mathrm{J}}}
\newcommand{\asunc}[3]{#1_{-#2}^{+#3}}
\newcommand{\asuncunit}[4]{#1_{-#2}^{+#3}\,\textrm{#4}}
\newcommand{\Av}{A_{\mathrm{V}}}
\newcommand{\bzk}{BzK }
\newcommand{\bvalue}{$b$-value}
\newcommand{\dlogJdlogM}{\mathrm{d}\,\log(j\,\mathrm{[kpc\, km\, /\, s]}) / \mathrm{d}\,\log(\Mstar\,[\Msun])}
\newcommand{\dlogRdM}{\mathrm{d}\,\log{r\,[\mathrm{kpc}]} / \mathrm{d}\,\log\Mstar\,[\Msol]}
\newcommand{\dlrdll}{\frac{\Delta\log \re}{\Delta\log \lambda}}
\newcommand{\dz}{ \mathrm{d}z}
\newcommand{\eagle}{EAGLE}
\newcommand{\eazy}{\textsc{EAZY}}
\newcommand{\estar}{\epsilon_{\mathrm{star}}}
\newcommand{\hlg}[1]{\colorbox{green}{#1}}
\newcommand{\hlr}[1]{\colorbox{red}{#1}}
\newcommand{\hly}[1]{\colorbox{yellow}{#1}}
\newcommand{\Hmag}{H_{\textrm{AB}}}
\newcommand{\hyperion}{\textsc{hyperion}}
\newcommand{\illustristng}{IllustrisTNG}
\newcommand{\Ks}{K_{\mathrm{s}}}
\newcommand{\lmass}{\log(\Mstar/\Msol)}
\newcommand{\lobs}{\lambda_{\mathrm{obs}}}
\newcommand{\lrho}{\log(\rho_1\,\textrm{kpc}^3/\Msol)}
\newcommand{\lsfr}{\log(\mathrm{SFR} / \Msol\,\mathrm{yr}^{-1})}
\newcommand{\lssfr}{\log(\mathrm{sSFR}\times \mathrm{yr})}
\newcommand{\magneticum}{Magneticum}
\newcommand{\Msol}{M_{\odot}}
\newcommand{\Mstar}{M_{\star}}
\newcommand{\Msun}{\Msol}
\newcommand{\mthree}{M3}
\newcommand{\omegabaryon}{\Omega_{\mathrm{B}}}
\newcommand{\omegalambda}{\Omega_{\Lambda}}
\newcommand{\omegamatter}{\Omega_{\mathrm{M}}}
\newcommand{\re}{r_{\mathrm{e}}}
\newcommand{\red}[1]{\textcolor{red}{#1}}
\newcommand{\rellip}{r^{\mathrm{e}}}
\newcommand{\rellipi}{\rellip_{i}}
\newcommand{\sersic}{Sérsic}
\newcommand{\sextractor}{\textsc{SExtractor}}
\newcommand{\skirt}{\textsc{skirt}}
\newcommand{\subfind}{\textsc{subfind}}
\newcommand{\tlb}{t_{\mathrm{LB}}}
\newcommand{\tng}{TNG}
\newcommand{\vc}{v_{1}}
\newcommand{\zphot}{z_{\textrm{phot}}}
\newcommand{\zsnap}{z_{\mathrm{snap}}}
\newcommand{\zspec}{z_{\textrm{spec}}}

\newcommand{\problem}[1]{%
  \rule{1\columnwidth}{0.25pt}\\
   \colorbox{yellow}{\parbox{\dimexpr\columnwidth-2\fboxsep-2\fboxrule\relax}{#1}}
}

\begin{acronym}
	\acro{agn}[AGN]{active galaxy nucleus}
	\acroplural{agn}[AGNs]{active galaxy nuclei}
	\acro{et}[ET]{early-type}
	\acro{lt}[LT]{late-type}
   \acro{fwhm}[FWHM]{full width at half maximum}
	\acro{hst}[HST]{Hubble Space Telescope}
	\acro{imf}[IMF]{initial mass function}
	\acro{irx}[IRX]{infrared excess}
	\acro{ms}[MS]{main-sequence}
	\acro{mtl}[$M/L$]{mass-to-light}
	\acro{nir}[NIR]{near-infrared}
	\acro{nmad}[NMAD]{normalized median absolute deviation}
	\acro{pbzk}[pBzK]{passive BzK}
	\acro{photoz}[photo-z]{photometric redshift}
	\acro{psb}[PSB]{post-starburst}
	\acro{psf}[PSF]{point spread function}
	\acro{rms}[RMS]{root mean square}
	\acro{sam}[SAM]{semi-analytic model}
	\acro{sed}[SED]{spectral energy distribution}
	\acro{sfh}[SFH]{star formation history}
	\acroplural{sfh}[SFHs]{star formation histories}
	\acro{sfr}[SFR]{star formation rate}
	\acro{smf}[SMF]{stellar mass function}
	\acro{snr}[SNR]{signal-to-noise ratio}
	\acro{ssfr}[sSFR]{specific star formation rate}
	\acro{ssp}[SSP]{simple stellar population}
	\acro{wfc3}[WFC3]{Wide Field Camera 3}
\end{acronym}
\acused{ssfr}

\label{firstpage}
\pagerange{\pageref{firstpage}--\pageref{lastpage}}
\maketitle

\begin{abstract}
We study stellar population and structural properties of massive $\lmass>11$ galaxies at $z\approx2.7$ in the Magneticum and IllustrisTNG hydrodynamical simulations and GAEA semi-analytic model.
We find stellar mass functions broadly consistent with observations, with no scarcity of massive, quiescent galaxies at $z\approx2.7$, but with a higher quiescent galaxy fraction at high masses in IllustrisTNG. Average ages of simulated quiescent galaxies are between $\approx0.8$ and $\SI{1.0}{Gyr}$, older by a factor $\approx2$ than observed in spectroscopically-confirmed quiescent galaxies at similar redshift. Besides being potentially indicative of limitations of simulations in reproducing observed star formation histories, this discrepancy may also reflect limitations in the estimation of observed ages. We investigate the purity of simulated UVJ rest-frame color-selected massive quiescent samples with photometric uncertainties typical of deep surveys (e.g., COSMOS). We find evidence for significant contamination (up to $\SI{60}{percent}$) by dusty star-forming galaxies in the UVJ region that is typically populated by older quiescent sources. Furthermore, the completeness of UVJ-selected quiescent samples at this redshift may be reduced by $\approx\SI{30}{percent}$ due to a high fraction of young quiescent galaxies not entering the UVJ quiescent region. Massive, quiescent galaxies in simulations have on average lower angular momenta and higher projected axis ratios and concentrations than star-forming counterparts. Average sizes of simulated quiescent galaxies are broadly consistent with observations within the uncertainties. The average size ratio of quiescent and star-forming galaxies in the probed mass range is formally consistent with observations, although this result is partly affected by poor statistics.
\end{abstract}




\section{Introduction}
\label{sec:intro}
In the local Universe spheroids contain about $\SI{75}{percent}$ of the total stellar mass \citep[including ellipticals, E0s and bulges in spirals,][]{Renzini_2006}. At the same time, about 10 times more quiescent than star-forming galaxies are found at $\lmass\gtrsim 11.5$ \citep[e.g.,][]{Baldry_2004}.
Indeed, there is a long-known strong correlation between morphology and stellar populations of nearby galaxies, where early-type, bulge-dominated systems have suppressed star-formation and generally host older stellar populations, whereas younger, actively star-forming galaxies typically have late-type, disk-like morphologies.
Morphologically early-type galaxies have higher central concentrations and lower apparent ellipticity. It is established that they are more compact than star-forming galaxies up to stellar masses $\lmass\approx 11$ \citep[e.g.,][]{Shen_2003, Guo_2009} and have lower specific angular momentum at fixed stellar mass \citep[e.g.,][]{Fall_1983specificAngularMomentum_Mass, Romanowsky_2012_AngularMomentum, Fall_2018_AngularMomentum}. \cite{Huang_2017_relation_size_angular_momentum_galaxy_ghalo} showed that there is a close connection between sizes and specific angular momenta of galaxies and those of their dark matter halos.

Studies of massive galaxies at higher redshift have revealed a much lower quiescent fraction, reduced by up to a factor of $\approx 10$ at $z\approx 3$ at the same stellar mass \citep[e.g.,][]{Muzzin_2013_SMF, Martis_2016_passive_fractions, Davidzon_2017_SMF_COSMOS2015}.
Morphological studies of massive quiescent galaxies at higher redshift have produced conflicting results. Some studies find quiescent galaxies with low \sersic\ indices and smaller axis ratios than at lower redshift \citep{VanDokkum_2008, McGrath_2008, Bundy_2010, VanDerWel_2011, Chang_2013, McLure_2013, Hsu_2014, Toft_2017, Bezanson_2018, Newman_2018}, suggesting that the strong correlation between morphology and broad stellar population properties is not as pronounced at earlier cosmic times. However, other studies \citep[e.g.,][]{Bell_2012, Lang_2014, Mowla_2019_b, Esdaile_2020} find evidence that the correlation already exists at high redshift.
The investigation of the relation between morphology and star formation properties is crucial to understanding quenching processes and is more generally a critical aspect of galaxy evolution.

The most famous morphological classification scheme for galaxies is the Hubble sequence \citep{Hubble_1926_fork}.
\cite{Sandage_1970_angular_momentum_hubble_sequence} suggested that the Hubble sequence is at fixed mass a sequence of increasing angular momentum.
The angular momentum of galaxies is obtained by primordial tidal torques \citep{Peebles_1969_OriginAngularMomentum}.
\cite{Fall_1983specificAngularMomentum_Mass} measured that the specific angular momentum at fixed stellar mass of nearby spiral galaxies is larger than for ellipticals. This result was later confirmed by \citet[][]{Romanowsky_2012_AngularMomentum}, \cite{Obreschkow_2014_mass_spin_morphology_spirals} and \cite{Cortese_2016_AngularMomentumMorphology}.
The lower specific angular momentum of elliptical galaxies has been investigated from a theoretical perspective in a number of studies \citep[e.g.,][]{Barnes_1987_angular_momentum_torques, Navarro_1994_dissipative_galaxy_formation_dynamics, Heyl_1996_structure_merger_remnants, Zavala_2008_angular_momentum_evolution_simulations, Lagos_2017_angular_momentum_evolution_EAGLE, Lagos_2018_mergers_impact_angular_momentum,
 Zoldan_2018_sizes_angular_momenta_GAEA, Rodriguez-Gomez_2022_AngularMomentumTNG}.
By measuring stellar kinematics with integral field spectroscopy it has been shown that in the local Universe the vast majority of early type galaxies are fast rotators and slow rotators dominate only on the high-mass end \citep[$\Mstar\gtrsim 2\times10^{11}\,\Msun$,][]{Emsellem_2011, Cappellari_2016}.
Another contribution to the evolution of the specific angular momentum with time comes from cosmic expansion for which \cite{Obreschkow_2015_angular_momentum_scaling} derive that for a spherical halo the specific angular momentum evolves with $(1+z)^{-1/2}$.

Concerning star formation properties, many studies over more than a decade have shown a close correlation between the \ac{sfr} and the stellar mass of star-forming galaxies, the so-called star-forming \ac{ms} that exists up to at least $z\approx 4$ \citep[e.g.,][]{Noeske_2007_main_sequence_general, Daddi_2007_main_sequence, Pannella_2009_star-formation_dust_main_sequence, Rodighiero_2011_main_sequence_z2_starbursts, Wuyts_2011_main_sequence, Whitaker_2012_main_sequence_z2.5, Schreiber_2015, Renzini_2015_main_sequence, 2020_Tacconi_starforming_interstellar_medium_over_time_review}.
It has been argued that galaxies with a steady balance between gas infall and \ac{sfr} lie on the \ac{ms} with a slope of 1, however, observations find lower values between $0.4$ and $1.0$, which is assumed to be a result of quenching processes and smaller cold gas fractions at higher stellar mass \citep[][]{Abramson_2014_main_sequence_slope, Pan_2017_main_sequence_slope, Pearson_2018_main_sequence_herschel}.
The scatter of the \ac{ms} is estimated to be approximately $\SI{0.3}{dex}$, independent of mass and at least up to $z\approx 3$ \citep[][]{Whitaker_2012_main_sequence_z2.5, Speagle_2014_main_sequence, Tomczak_2016_main_sequence_since_z4, Pearson_2018_main_sequence_herschel}. This is believed to be a result of a cyclic process in which galaxies achieve a higher \ac{sfr} as a consequence of compactification of the gas in their center that triggers star-formation. This is followed by depletion of the gas and reduced \ac{sfr} until new infalling gas fuels the star-formation again. When the replenishment time becomes longer than the depletion time the galaxy becomes quiescent \citep[][]{Tacchella_2016_main_sequence_scatter_quenching_model, Pearson_2018_main_sequence_herschel}.
The normalization of the \ac{ms} increases at higher redshift \citep[e.g.,][]{Speagle_2014_main_sequence, Johnston_2015_SFR_stellar_mass_since_z3, Schreiber_2015, Tomczak_2016_main_sequence_since_z4} as expected due to the higher amount of cold gas available for star formation \citep[][]{Tacconi_2008_cold_gas_fraction_redshift, Dunne_2011_cold_gas_fraction_redshift, Genzel_2015_gas_fraction_time, Scoville_2016_dust_gas_star-formation_redshift, Pearson_2018_main_sequence_herschel, Tacconi_2020_SF_ISM_across_time}.

While at low redshift a specific star formation rate ($\ac{ssfr} = \mathrm{SFR} / \Mstar$) threshold of $\lssfr=-11$ is often used to classify galaxies as star-forming or quiescent \citep[e.g.,][]{Brinchmann_2004, Fontanot_2009_sSFR_cut}, the evolving \ac{ms} can be used at higher redshift to define a threshold for quiescence to account for the generally higher \ac{sfr} at earlier cosmic times.
However, since robust \ac{sfr} estimates require unbiased \ac{sfr} tracers rarely available over large, mass-limited samples, especially at high redshift the quiescent vs. star-forming classification of galaxies is often done by means of observed or restframe colors that are able to separate quiescent and star-forming galaxies accounting for the impact of dust attenuation \citep[e.g.,][]{Daddi_2004_BzK, Williams_2009, Arnouts_2013_NUVrK, Ilbert_2013}. At higher redshift increasing photometric uncertainties lead to larger uncertainties on the classification and to potentially strong contamination of quiescent samples by star-forming sources that may bias derived properties of quiescent galaxies. Additionally, quiescent samples at high redshift contain a larger fraction of relatively young sources that would be classified as post-starburst galaxies at lower redshift \citep[e.g.,][]{Whitaker_2012, Marchesini_2014, Merlin_2018, Maltby_2018, Lustig_2021, Deugenio_2020_letter}. As shown in some studies \citep[e.g.,][]{Merlin_2018, Schreiber_2018_NIR_spec_3z4, Forrest_2020_survey_rapid_SF_and_quenching_early_universe} a potentially significant population of such young quiescent galaxies might be missed by using standard photometric classification criteria.

The comparison of galaxy properties in observations and simulations allows a more detailed investigation and improvement of our theoretical understanding of the evolution of galaxies, at the same time enhancing our capabilities of interpreting observations disentangling different aspects.
Indeed, the discovery of massive quiescent galaxies at high redshift has shown a tension with simulations that predicted too low quiescent fractions \citep[e.g.,][]{Straatman_2014_quiescent_galaxy_population_z4, Glazebrook_2017, Schreiber_2018_NIR_spec_3z4, Merlin_2019_quiescent_galaxies_at_dawn_of_universe, Guarnieri_2019_z4_galaxy_candidates, Alcalde-Pampliega_2019_zlarger3_galaxies, Forrest_2020_one_massive_galaxy_z3.5}, and recent James Webb Space Telescope observations have revealed even higher number densities of quiescent galaxies at high redshift than found before \citep[][]{Carnall_2022_JWST_Q_galaxies}. In trying to mitigate discrepancies in quiescent fractions at high redshift from observations and simulations, an important role is played by better physical descriptions of \acp{agn} which can quench star formation and eject gas from the host galaxy. \ac{agn} feedback could increase the fraction of quiescent galaxies to approach agreement with observations and is potentially also an important contribution for morphological evolution \citep[e.g.,][]{McCarthy_2011_AGN_feedback_overcooling_solution, Dubois_2016_morphology_AGN_feedback, Remus_2017_morphology_feedback, Kaviraj_2017_Horizon_AGN_noAGN, Weinberger_2018_AGN_feedback_effect_TNG, Fontanot_2020_AGN_in_simulations}.
In addition, the number of resolution elements in hydrodynamical simulations has increased significantly in recent years, allowing simulations with larger volume and higher resolution to investigate spatially resolved properties of galaxies \citep[e.g.,][]{Genel_2014_galaxy_populations_cosmic_time_resolution_evolution}.

On the observational side to reduce model ambiguities and provide conclusive evidence of the high redshift and quiescence of very distant, candidate quiescent galaxies, spectroscopic observations are necessary: current spectroscopic confirmation of quiescent galaxies reaches out to $z\approx 4$ \citep{Glazebrook_2004, Cimatti_2004, Kriek_2006, Gobat_2012, Onodera_2012, Onodera_2015, Newman_2015, Marsan_2015, Hill_2016, Glazebrook_2017, Marsan_2017, Gobat_2017, Newman_2018, Schreiber_2018_NIR_spec_3z4, Tanaka_2019, Forrest_2020_one_massive_galaxy_z3.5, Forrest_2020_survey_rapid_SF_and_quenching_early_universe, Valentino_2020_Q_1.5years_after_binbang_and_progenitors, Esdaile_2020}.
In \cite{Deugenio_2020_letter, DEugenio_2020_paper} and \cite{Lustig_2021} we presented an analysis of stellar population and structural properties of one of the first sizeable samples of $z\approx 3$ spectroscopically confirmed quiescent galaxies, with the homogeneous investigation of 10 massive sources at $2.4<z<3.2$, extending previous work on spectroscopically confirmed samples \citep{Cimatti_2008, VanDokkum_2008, Belli_2017, Stockmann_2020} to higher redshift.

In this paper we compare observational findings on morphology and stellar population properties of massive quiescent galaxies at $z\approx 3$, with current expectations from theoretical models of galaxy evolution in a cosmological context. We focus in particular on hydro-dynamical simulations, and specifically \magneticum\ and \illustristng\ (hereafter \tng), that are able to provide spatially resolved galaxy properties to attempt a more direct comparison with observational results (Sections~\ref{sec:magneticum},~\ref{sec:tng}). For context and broader completeness in our analysis we also consider, for some aspects discussed in the following, a comparison with the semi-analytical model GAEA (Sections~\ref{sec:gaea}), which does not provide resolved galaxy properties but helps improving our general picture of the comparison between observations and theoretical expectations by providing a large volume (particularly important for intrinsically rare sources as those investigated here) and a different modelling approach with respect to the hydro-dynamical counterparts.
We also take advantage of the simulation framework to investigate the performance, limitations and biases of the routinely adopted photometric selection of quiescent galaxies at this redshift.

In Section~\ref{sec:data} we introduce the simulations used in this paper, and present the investigated simulated galaxy samples and main observational studies used here for comparison. In Section~\ref{sec:SMF} we present \acp{smf} at $z\approx3$ as predicted in the simulations in the context of observational determination and in Section \ref{sec:stellarages} we discuss stellar ages of galaxies. In Section~\ref{sec:uvj} we discuss photometric selection of quiescent galaxies at high redshift and in Section~\ref{sec:morphology} we present a morphological analysis of simulated galaxies. We summarize our findings and conclusions in Section~\ref{sec:conclusion}. Despite some small differences the cosmology adopted in all the different parts contributing to this analysis is a flat $\Lambda$CDM cosmology with $\omegamatter\approx0.3$ and $h\approx0.7$ (see more specific details in Section~\ref{sec:sampleselection} and Table~\ref{tab:simulations_general}). We assume a \cite{Chabrier2003} \ac{imf}. Magnitudes are given in the AB system.

\section{High-redshift quiescent galaxies in simulations and observations}
\label{sec:data}

In this work we compare observational results on very distant quiescent galaxies with theoretical counterparts from the IllustrisTNG and Magneticum suites of hydrodynamical simulations.
We focus on simulations with a large enough volume to host a large enough number of massive galaxies at $z\approx2.7$ and a high enough resolution for a morphological analysis. As mentioned in Section~\ref{sec:intro}, we also consider, wherever possible and appropriate, a comparison with results from the semi-analytic model GAEA. In the following we briefly introduce the simulations used here.
Some of their basic properties are listed in Table~\ref{tab:simulations_general} and we refer to the relevant publications listed below for a throughout description and specific details.

\subsection{Simulation data}
\subsubsection{IllustrisTNG}
\label{sec:tng}
IllustrisTNG\footnote{www.tng-project.org} \citep{Springel_2018_firstresultstng, Nelson_2018_firstresultstng, Marinacci_2018_firstresultstng, Pillepich_2018_firstresultstng, Naiman_2018_firstresultstng} is a set of magneto-hydrodynamical cosmological simulations with different physical sizes and mass resolutions. In this work we use the public data release \citep[][]{Nelson_2019} of the simulations TNG100 and TNG300, with box sizes of 100 and $\SI{300}{Mpc}$, at the highest available resolution. Baryonic particle masses in TNG300 and TNG100 are $1.1\times10^7\, \Msun$ and $1.4\times 10^6\,\Msun$, respectively. A \citet{Planck_2016_cosmological_parameters} cosmology is used in both simulations with $\mathrm{h}=0.677, \omegalambda=0.691, \omegamatter=0.309$ and $\omegabaryon=0.049$.
The comoving softening length for stellar particles is $\SI{2.0}{kpc/h}$ in TNG300 and $\SI{1.0}{kpc/h}$ in TNG100.
Details of the galaxy formation models can be found in \cite{Weinberger_2017_galaxy_fromation_methods} and \cite{Pillepich_2018_galaxy_fromation_methods}. 

\subsubsection{Magneticum Pathfinder}
\label{sec:magneticum}
The Magneticum Pathfinder\footnote{www.magneticum.org} simulations are a suite of fully hydrodynamical cosmological simulations, covering different box volumes and resolutions \citep[see][]{Hirschmann_2014_magneticum_black_hole_growth, Teklu_2015_magneticum_angular_momentum_galaxy_dynamics}.
Here, we use the Magneticum Pathfinder simulation box 3 (hereafter \mthree) with the highest available resolution \citep[][]{Steinborn_2016_box3}. The simulated box has a comoving side length of $\SI{128}{Mpc/h}$ containing initially $2\times 1536^3$ particles (dark matter and gas) with a resolution of $m_{\mathrm{DM}}=3.6\times 10^7\,\Msun$ and $m_{\mathrm{gas}}=7.3\times 10^6\,\Msun$ with each gas particle spawning up to 4 stellar particles, resulting in an average particle mass of $\approx 1.8\times10^6\,\Msun$. The comoving softening length for stellar particles is $\estar=\SI{0.7}{kpc/h}$.
A WMAP7 $\Lambda$CDM cosmology \citep{Komatsu_2011_wmap7} is used for the simulation with $\sigma_8= 0.809, \mathrm{h}=0.704, \omegalambda=0.728, \omegamatter=0.272$ and $\omegabaryon= 0.045$. \mthree\ provides a refined black hole accretion and \ac{agn} feedback model with respect to the implementation from \cite{Springel_2005_GADGET2} that results in a better agreement with the observed black hole mass - stellar mass relation due to faster black hole growth at higher redshifts. For further details we refer to \cite{Steinborn_2015}.

\subsubsection{GAEA}
\label{sec:gaea}
The GAlaxy Evolution and Assembly (GAEA) model \citep{delucia2014, hirschmann2016, xie2017, DeLucia_2019_SF_regulation_in_satelites, Fontanot_2020_AGN_in_simulations} is a state-of-the-art \ac{sam}.
In contrast to hydro-dynamical simulations, \acp{sam} follow the evolution of galaxies by modelling the key physical processes acting on the baryonic component by means of physically or empirically motivated prescriptions. These theoretical tools are usually built upon a statistical description for dark matter halo merger trees, that can be derived either from dark matter only simulations or from analytical perspective.
As a result, and of immediate relevance to this work, while hydrodynamical simulations directly predict the 3D spatial distribution (and thermodynamical properties) of baryonic components, in \acp{sam} the internal structure of both galaxies and dark matter halos (including their thermal and kinematical properties) is not resolved, but derived statistically.
The coupling with large dark matter only simulations allows access to a very large dynamic range in mass and spatial resolution. In addition, the limited computational costs allow an efficient investigation of the parameter space and influence of different physical assumptions, as well as access to large simulated volumes which is particularly critical for intrinsically rare sources, as relevant in this study. A number of studies have presented dedicated comparisons between results from semi-analytical and hydro-dynamical models \citep[e.g.,][]{guo2016, knebe2018, asquith2018, wang2018, wang2019, ayromlou2021}, with the broad intent of investigating the impact of the prescriptions adopted in both.

In this work, as a complementary comparison between observation and theoretical models, we consider predictions from the GAEA-F06 implementation described in \citet{Fontanot_2020_AGN_in_simulations}. This is based on merger trees from the Millennium Simulation \citep{springel2005}, a numerical realization consisting of 2160$^3$ particles with a box size of 500~Mpc~$h^{-1}$, assuming a WMAP1 cosmology\footnote{The mismatch of these parameters with more recent measurements from Planck and WMAP \citep{planckcoll2016, bennett2013} does not significantly affect model predictions once the model parameters are adjusted to reproduce observables in the local Universe \citep[e.g.,][]{Guo_2013_Galaxy_formation_WMAP1_vs_7}.} with $\Omega_{\Lambda} = 0.75$, $\omegamatter=0.25$, $\omegabaryon = 0.045$,  $\sigma_8 = 0.9$, $h=0.73$.
The resolution of the simulation is adequate to resolve galaxies down to $\approx 10^9\,\Msun$.
The model accounts for a range of key physical processes including cooling and heating of baryonic gas, star formation, gas accretion onto SMBHs, stellar and AGN feedback, chemical enrichment, bulge formation during mergers and driven by disc instability (full details can be found in the GAEA references above).

Although GAEA may provide some information about the galaxy structural properties \citep[][]{xie2017} and angular momentum \citep[][]{Zoldan_2018_sizes_angular_momenta_GAEA}, in this work we limit this part of the comparison to hydrodynamical simulations because our analysis, by analogy with observational studies, uses spatially resolved information for simulated galaxies, making the comparison with the global value predicted by semi-analytic models non-trivial.

\subsection{Observational studies used for comparison}
\label{sec:observational_studies}
In this work, theoretical predictions are compared to a range of observational properties of distant quiescent galaxies drawn from several studies. We focus in particular on the number density of quiescent galaxies and quiescent fractions at $z\approx3$ as estimated in deep fields from \cite{Ilbert_2013}, \cite{Muzzin_2013_SMF}, \cite{Martis_2016_passive_fractions}, \cite{Davidzon_2017_SMF_COSMOS2015} and \cite{Sherman_2020_passive_fractions}, and on structural and stellar population properties both from statistical photometric samples from deep-field photometric studies \citep[][]{VanDerWel2014, LaigleCOSMOS2016} and more specifically from our dedicated follow-up of a spectroscopically confirmed sample of massive (median stellar mass $\lmass\approx 11.2$) quiescent galaxies at $z\approx 3$ \citep[][]{Deugenio_2020_letter, DEugenio_2020_paper, Lustig_2021}.

Indeed, this work is partly intended as a simulation analysis counterpart to our previous observational studies of this sample. Therefore, we use results from \cite{Deugenio_2020_letter, DEugenio_2020_paper} and \cite{Lustig_2021} as a preferential observational counterpart in the following, specifically commenting on relevant results from other studies as needed. We summarise here the main aspects of these observations.
The sample of ten galaxies was targeted with the \acl{wfc3} on the Hubble Space Telescope with the F160W filter (H band) and the G141 grism. Targets were selected as $z>2.5$ quiescent galaxy candidates based on photometric \citep[BzK and UVJ,][]{Daddi_2004_BzK, Williams_2009} classification, excluding potential star-forming contaminants and focusing on the brightest sources with $H\lesssim 22$ for observational reasons \citep[for full details on sample selection see][]{DEugenio_2020_paper, Lustig_2021}. As a results, this sample is made of very massive galaxies at this redshift, with nine out of ten sources having estimated stellar masses $\lmass >11$. \citet[][]{Deugenio_2020_letter, DEugenio_2020_paper} measured spectroscopic redshifts from grism data between $z=2.4$ and $z=3.2$ (median $z=2.7$) and confirmed quiescence for all targets, with estimated stellar ages between $300$ and $\SI{800}{Myr}$, consistent with an average formation redshift of $z\approx 3.5$.
In \cite{Lustig_2021} we analysed the morphology of galaxies in this sample and found high \sersic\ indices and axis ratios (medians $\approx 4.5$ and $0.73$, respectively), pointing towards a largely bulge dominated population among quiescent galaxies already at $z\approx 3$. For further details we refer to the aforementioned papers.

\subsection{Sample selection of simulated galaxies and definitions}
\label{sec:sampleselection}

We select simulated galaxies in the available simulation snapshots closest to the median redshift of our main observed comparison sample \citep[][]{Deugenio_2020_letter, DEugenio_2020_paper, Lustig_2021}, which is $z=2.83$ in GAEA, $z=2.73$ in both TNG simulations and $z=2.79$ in \mthree. To compare with the massive galaxies in our observed sample (median stellar mass $\lmass\approx 11.2$) we focus on massive galaxies with $\lmass\geq11.0$. For the hydrodynamical simulations, halo structures are identified with \subfind\ \citep{Springel_2001_subfind, Dolag_2009_subfind}, and we initially select all galaxies above the $\lmass=11$ mass threshold considering all gravitationally bound particles according to \subfind. This yields a parent sample of 196 galaxies in \mthree, 1077 in TNG300 and 78 in TNG100.
In GAEA we find 9339 galaxies above this mass threshold.
For all simulations, the selected galaxies are classified as quiescent or star-forming in the following, as described in Section~\ref{sec:passivefractions}. A comparison with photometric classification criteria as used in observations is discussed in Section~\ref{sec:uvj}.

The comparison between stellar masses from observations and simulations is not trivial. Statistical and systematic uncertainties on stellar masses from \ac{sed} fitting are estimated to be a factor of $\approx 2$ \citep[e.g.,][]{Maraston_2006, Longhetti_2009, Muzzin_2009, Conroy_2013, Pacifici_2015}. On the simulation side, stellar masses assigned to simulated galaxies rely 
anyway on assumptions (in particular concerning the \ac{imf}, but also more generally the modeling of stellar populations), and are thus not necessarily directly comparable to the observational estimates. As usual in this kind of study, in the following we do compare nonetheless the stellar masses of simulated and observed galaxies, in the attempt to select comparable samples whose properties can be contrasted. We caution that any systematics in the stellar mass scale between observations and simulations would affect such comparisons. Furthermore, from a more practical point of view, because of the limited \ac{snr} in observations, the inclusion of all bound stellar particles is not comparable to what occurs in the analysis of observed galaxies \cite[see also e.g.,][]{Pillepich_2018_firstresultstng, Genel_2018_TNGSizeEvolution, Donnari_2019, Donnari_2021_comparison_observations}. For a more realistic comparison we define different subsamples for the hydrodynamical simulations, defined by including only bound particles within a certain distance from the center of the host galaxy. Because of the complex behaviour of the \ac{snr} of observed galaxies and of the corrections applied to estimate the total galaxy light, it is not trivial to find an aperture that matches those used in observations. We therefore use a range of 2-dimensional apertures (by projecting the simulated galaxies along random lines of sight) to estimate the systematic uncertainties implied by the aperture choice and the impact on our analysis. We carry out the analyses in the following for four different aperture choices (sizes refer to physical kpc): $\SI{30}{kpc}$ apertures (corresponding to $\approx\SI{3.7}{arcsec}$ at $z=2.7$) around both the center of mass and center of light (in the observed H-band, see Section~\ref{sec:luminosities} for the calculation of the adopted synthetic luminosities for simulated galaxies), and apertures equal to 2 times the half-mass and half-light radii. The average offset between the mass and light-weighted center is less than $\SI{0.2}{kpc}$ in the considered simulations, the choice between the two has no impact on our results. The half-light and half-mass radii are in the order of $1-\SI{4}{kpc}$ for the kind of galaxies studied here (see Section~\ref{sec:morphology}). The large range of sizes covered by these apertures brackets typical sizes of apertures in observations \citep[see also e.g.,][]{Donnari_2019, Donnari_2021_comparison_observations}.
We therefore obtain 4 different subsamples of massive galaxies by applying again the mass cut of $\lmass\geq11$ to the same simulated parent galaxy sample by accounting for particles within the 4 different apertures considered.

To calculate the centers (of mass and light) of simulated galaxies for defining apertures, we project the particle positions along random lines of sight and iteratively calculate the average position of particles weighted by either their stellar mass or H-band luminosity, excluding particles at $>\SI{2}{kpc}$ from the center\footnote{Visual inspection confirms that a threshold of $\SI{2}{kpc}$ gives the best estimate of the halo centers for most of the subhalos.} until convergence. Identifying as the galaxy center the position of the local (mass or light) density maximum within $\SI{10}{kpc}$ of the above-defined center - possibly a better definition for a minor fraction of asymmetric galaxies - results in an average shift of the assumed center of the aperture by $\SI{0.16}{kpc}$, and has no impact on the results presented here.

As discussed in Section~\ref{sec:passivefractions} we present by default our results obtained with apertures of $\SI{30}{kpc}$ around the center of mass and commenting the impact of the aperture choice where needed. The sample sizes with this aperture are 166, 993 and 73, for \mthree, TNG300 and TNG100, respectively. The median stellar mass of these samples is $\lmass=11.1$ in all simulations.

\subsection{Estimation of luminosities for simulated galaxies}
\label{sec:luminosities}
For some analyses we use restframe U, V, J and observed H band luminosities.
In particular, we compute luminosities for galaxies in hydrodynamical simulations by assigning to each stellar particle a \ac{sed} corresponding to its age using \cite{Bruzual2003} stellar population synthesis models with a \cite{Chabrier2003} \ac{imf}, linearly interpolating the models to match the metallicity of the given particle. We then calculate global \acp{sed} as the sum of the individual particle \acp{sed} within the considered apertures.

To properly relate these stellar \acp{sed} to observed photometry of real galaxies, we need to account for the impact of dust attenuation, for which we adopt an empirical approach.
For star-forming galaxies we use a \cite{Calzetti2001} dust attenuation law with stellar mass dependent attenuation strength from 1) \citet[][]{McLure_2018_dust_attenuation}, who find $A_{1600}=2.293+1.160X+0.256X^{2}+0.209X^3$, where $A_{1600}$ is the attenuation at $\SI{1600}{\angstrom}$ and $X=\log(\Mstar/10^{10}\Msun)$, resulting in $\Av=\SI{1.6}{mag}$ for a $\lmass=11$ galaxy, with a scatter of $\SI{0.4}{mag}$ and 2) \citet[][]{Pannella_2015_dust_extinction} who find $A_{\mathrm{UV}}=1.6\times(\lmass+0.26)-13.5$, giving $\Av=\SI{1.6}{mag}$ for a $\lmass=11$ galaxy, with a scatter of $\SI{0.2}{mag}$.
Given the similarity of the attenuation estimated from the work of \cite{McLure_2018_dust_attenuation} and \cite{Pannella_2015_dust_extinction}, for the sake of simplicity results presented in the following for star-forming galaxies are quoted only assuming $\Av$ from \cite{Pannella_2015_dust_extinction}. Using $\Av$ from \cite{McLure_2018_dust_attenuation} does not affect our results.

For quiescent galaxies at redshifts probed here, actual constraints on dust attenuation are much looser than for star-forming counterparts. With the empirical approach described below, we try to bracket a plausible range based on available observations of quiescent galaxies at $z\approx 3$, as well as of low-redshift counterparts. We start by adopting results from \cite{DEugenio_2020_paper} for our sample of 10 quiescent galaxies at $2.4<z<3.2$, finding $0.1 < \Av < 1.6$ with a median $\Av = 0.5$ \citep[other spectroscopic studies at similar redshift find $\Av$ in good agreement, e.g.,][]{Schreiber_2018_NIR_spec_3z4, Valentino_2020_Q_1.5years_after_binbang_and_progenitors, Esdaile_2020}.
For each simulated quiescent galaxy we randomly choose an $\Av$ from this distribution and consequently apply dust reddening to its \ac{sed} assuming a \cite{Calzetti2001} attenuation law modified by a power law with slope $\delta=-0.4$ \citep{Noll_2009_dust_attenuation, Salim_2018_dust_attenuation}.
However, since all galaxies in the \cite{DEugenio_2020_paper} sample are relatively young with $t_{50}<\SI{0.8}{Gyr}$ (where $t_{50}$ is the lookback time when half of the stellar mass of the galaxy was formed, see Section~\ref{sec:stellarages}) and a median $t_{50}=\asuncunit{0.45}{0.10}{0.05}{Gyr}$, adopting the attenuation values estimated by \cite{DEugenio_2020_paper} might overestimate the attenuation for older quiescent galaxies. To gauge the impact of this effect on our results, we also calculate luminosities by applying dust attenuation with \cite{DEugenio_2020_paper} $\Av$ only to quiescent galaxies with $t_{50}<\SI{0.8}{Gyr}$, while for older sources, due to the lack of measurements at this redshift, we adopt a \cite{Calzetti2001} dust attenuation law with an $\Av$ typical of quiescent galaxies with stellar ages $1-\SI{2}{Gyr}$ - as appropriate for old populations at $z\approx 3$ - from lower redshift work \citep{Gonzalez-Delgado_2015_dust}. We therefore assume for simulated quiescent galaxies with $t_{50}>\SI{0.8}{Gyr}$ a randomly selected $\Av$ from a Gaussian distribution with mean $\SI{0.25}{mag}$ and standard deviation $\SI{0.15}{mag}$.
As this estimate is based on low-redshift quiescent sources (likely affected by, if anything, less dust attenuation than $z\sim3$ counterparts of similar mass and age) we assume that our two empirical approaches should bracket the actual dust attenuation affecting $z\sim3$ quiescent galaxies in the probed mass range. In the following we quote as a default results assuming this latter prescription for dust attenuation of quiescent galaxies, and discuss as needed the dependence of the results on this choice.

For some purposes of our analysis we need to mimic photometric uncertainties. We quantify these based on the uncertainties (as a function of magnitude) on the photometry of galaxies at $2.5<z<3$ in the COSMOS2015 \citep{LaigleCOSMOS2016} catalog, whose data have been used for several observational work on high-redshift quiescent galaxies and in particular for studies we directly compare with in the following.

\subsection{Star-forming main sequence and selection of quiescent galaxies in simulations}
\label{sec:quiescent_selection}
Galaxies are often classified - and especially so at high redshift - as star-forming or quiescent based on their position in rest-frame or observed color diagrams \citep[e.g.,][]{Daddi_2004_BzK, Williams_2009, Arnouts_2013_NUVrK, Ilbert_2013}.
In simulations, different \ac{ssfr} criteria are used to separate quiescent and star-forming galaxies throughout the literature. If a \ac{ms} of star-forming galaxies \citep[e.g.,][]{Noeske_2007_main_sequence_general} can be identified in the simulated sample, galaxies with a SFR significantly lower than the \ac{ms} can be defined as quiescent \citep[e.g.,][]{Genel_2018_TNGSizeEvolution, Donnari_2019, Donnari_2021_comparison_observations}.
Alternatively, an absolute \ac{ssfr} threshold, typically depending on redshift \citep[e.g.,][]{Franx_2008}, can be defined to separate star-forming and quiescent sources. In this section we compare the impact of such different criteria on the selection of high-redshift massive quiescent galaxies in the simulations investigated here.

We estimate the \ac{ms} following \cite{Donnari_2019}; we bin galaxies in $\SI{0.2}{dex}$ stellar mass bins and iteratively calculate the median \ac{sfr} in each bin, excluding galaxies with a \ac{sfr} more than $\SI{1}{dex}$ below the median until convergence. The uncertainty on the derived \ac{ms} is estimated by bootstrapping over 1000 samples.

\begin{figure*}
	\includegraphics[width=\textwidth]{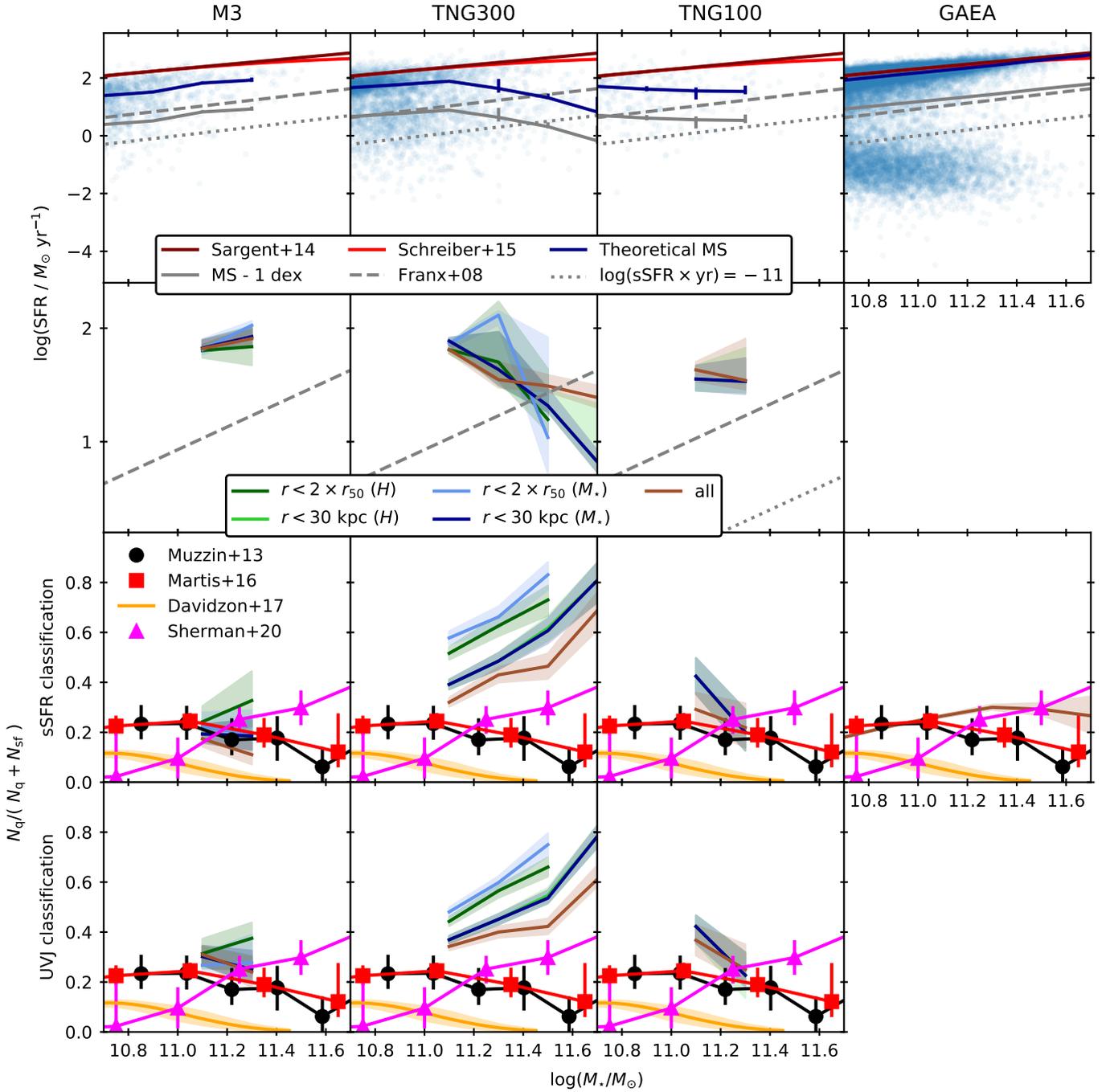}
	 \caption{star formation rates (top two rows) and quiescent galaxy fraction (bottom rows) as a function of stellar mass for galaxies at $z\approx 2.7$ in the studied simulations. All relations are estimated in $\SI{0.2}{dex}$ bins of stellar mass and only bins with at least 15 galaxies are shown.
	 {\it First row}: the \ac{ms} of star-forming galaxies (blue solid line) as determined in the \mthree, TNG300, TNG100 and GAEA simulations, as indicated at $\lmass>10.6$ (stellar masses and \acp{sfr} for \mthree\ and TNG galaxies are measured in $\SI{30}{kpc}$ apertures, see Section~\ref{sec:sampleselection}). Blue circles show individual galaxies (19 galaxies in \mthree\ and 7 galaxies in TNG300 have $\mathrm{SFR}=0\,\Msun\mathrm{yr}^{-1}$ and are not shown). The gray solid line shows the \ac{ms} offset to lower \ac{sfr} by \SI{1}{dex}. The dashed gray line shows the \protect\cite{Franx_2008} criterion for quiescence. Dotted gray line shows a $\lssfr=-11$. Observational determinations of the \ac{ms} at the same redshift are shown, as indicated. {\it Second row:} the dependence of the estimated \ac{ms} at high masses on the aperture used to compute galaxy stellar mass and \ac{sfr} in hydrodynamical simulations (see Section~\ref{sec:sampleselection}). 
	 Dark green (light blue) lines refer to apertures of $2\times$ the half light (half mass) radius. Light green (dark blue) lines refer to $\SI{30}{kpc}$ apertures around the center of light (mass). Brown lines refer to all particles associated with the galaxy. The light green and dark blue lines overlap in all simulations. Dashed and dotted lines are the same as in top panels.
	 {\it Third row:} the estimated quiescent fraction in the simulations as a function of stellar mass, and its dependence on the adopted aperture as indicated. Quiescent fractions shown here are defined based on the \protect\cite{Franx_2008} criterion for quiescence. Observational estimates of the quiescent fraction at the same redshift are shown, as indicated (see Section~\ref{sec:passivefractions} for full details, including definition of quiescent galaxy samples). {\it Fourth row:} Same as third row, but quiescent fractions in the simulations are estimated from UVJ color diagrams (see Section~\ref{sec:uvj}).
	}
	 \label{fig:mainsequence}
\end{figure*}

In Figure~\ref{fig:mainsequence} we show the estimated \ac{ms} of star-forming galaxies at $z\approx2.7$ for the simulations used in this paper and for different apertures (see Section~\ref{sec:sampleselection}). In this figure we include for clarity lower mass galaxies down to $\lmass=10.6$.
Estimates for different apertures in hydrodynamical simulations are consistent within the uncertainties, as shown in the figure.
The \ac{ms} in \mthree\ rises linearly from $\lsfr\approx 1.4$ at $\lmass\approx 10.6$ to $\lsfr\approx 2.1$ at $\lmass\approx 11.5$. At low masses the main sequence in \tng 300 is about $\SI{0.2}{dex}$ higher than in \mthree; it rises with a similar slope as in \mthree\ but bends at $\lmass\approx11.1$ ($\approx 11.3$ for the two smallest apertures) reaching down to $\lsfr\approx 1.3$ at $\lmass=11.5$. This is because a large fraction of massive galaxies in TNG300 is already quenched 
\citep[by AGN feedback, as discussed in][]{Weinberger_2017_galaxy_fromation_methods, Weinberger_2018_AGN_feedback_effect_TNG, Nelson_2018_firstresultstng, Zinger_2020_TNG_BH_feedback_galaxies, Donnari_2021_quenched_fractions_AGN_environment_preprocessing}
and no sequence of star-forming galaxies can be clearly identified. The \ac{ms} in TNG100 is flat over the full mass range considered here with a value of $\lsfr\approx 1.6$, however, we note the lack of very high-mass galaxies in this simulated volume.
The median \ac{sfr} of star-forming galaxies in GAEA is slightly higher than in hydrodynamical simulations and the \ac{ms} rises linearly from $\lsfr=1.9$ at $\lmass=10.7$ to $\lsfr=2.8$ at $\lmass=11.7$ \citep[see also][]{Wang_2019_main_sequence_GAEA}.

\ac{ms} determinations from observations at similar redshift \citep[e.g.,][see brown and red lines in the upper panels in Figure~\ref{fig:mainsequence}]{Sargent_2014_main_sequence, Schreiber_2015} are in very good agreement with predictions from GAEA and have slopes in overall good agreement with predictions from M3 and TNG300 at lower masses. However, observed \acp{ms} show higher \acp{sfr} by $\approx\SI{0.6}{dex}$ with respect to both M3 and TNG300 simulations. 
In this respect, we note that some studies \citep[e.g.,][]{Nelson_E_2021_SFR_obs_vs_sim_offset, Leja_2022_SFR_offset_sims_obs} suggested that an overestimation of the \acp{sfr} of observed galaxies might in fact explain the offset between the \ac{ms} in observations and simulations in previous studies, and this could be the case also at $z\approx 3$.

We stress that the estimated main-sequence location depends on the adopted definition of star-forming galaxies, and thus on the adopted approach. In particular, estimating the main sequence in simulations as described above is clearly affected by the level of arbitrariness in identifying the star-forming population, especially in those cases where the quiescent galaxy fraction is substantial, and/or there is no evident bimodality in the sSFR distribution of galaxies \footnote{We note that the existence of a bimodality in the actual \ac{sfr} distribution of galaxies is debated \citep[e.g.,][]{Feldmann_2017_are_SFRs_bimodal}. We only mention it here in relation to the impact of the chosen quiescence definition in the identification of the quiescent galaxy samples in simulations.}.
This is particularly the case for TNG, where indeed the very strong bending of the main-sequence in the high mass range probed here, as visible in Figure~\ref{fig:mainsequence}, has been extensively discussed in terms of adopted definitions and approaches in \cite{Donnari_2019}.

An often used \ac{sfr} threshold for defining quiescence is $\SI{1}{dex}$ below the main sequence \citep[see e.g.,][]{Morselli_2019_mass_sfr_along_main_sequence, Donnari_2019, Sherman_2020_passive_fractions}
which we show for measurements in an aperture of $\SI{30}{kpc}$ in Figure \ref{fig:mainsequence}. This definition depends directly on the way the \ac{ms} itself is estimated, also in relation to the specific characteristics of the given galaxy sample, and is thus affected by the arbitrariness in the \ac{ms} definition discussed above. For comparison, we thus also show in Figure \ref{fig:mainsequence} the \cite{Franx_2008} criterion to define quiescence, $\mathrm{sSFR} < 0.3\times H(z)$ \citep[see also][in particular on a study in Magneticum]{Lotz_2021_magneticum_post_starburst}, that is $\lssfr\approx -10.0$ at $z=2.75$. We stress that, although the \cite{Franx_2008} criterion does not depend on the approach we adopt to define the \ac{ms}, and is thus less arbitrary in that respect, defining quiescence with any given fixed \ac{ssfr} threshold still remains to some extent arbitrary in relation to the \ac{ssfr} distribution of the considered galaxy population. Indeed we note that with the available statistics we cannot identify any bimodality in the \ac{ssfr} distribution of $\lmass>11$ galaxies in M3 and TNG100. We do identify a clear bimodality in GAEA (as visible in Figure~\ref{fig:mainsequence}), with both the $\mathrm{MS}-\SI{1}{dex}$ and \cite{Franx_2008} criteria approximately falling in the gap between the quiescent and star-forming populations. For TNG300, we can see a hint of bimodality in the \ac{ssfr} distribution, with the \cite{Franx_2008} threshold falling to slightly lower \acp{ssfr} with respect to the gap (and then consequently the $\mathrm{MS}-\SI{1}{dex}$ threshold may fall within the low-\ac{ssfr} population at higher masses, due to the MS bending shown in Figure~\ref{fig:mainsequence}).

As the figure shows, in \mthree\ and GAEA the \ac{ms}-based and \cite{Franx_2008} criteria give relatively similar \ac{ssfr} thresholds over the full mass range, with the \ac{ms}-based threshold being lower (higher) by $\SI{0.3}{dex}$ ($\SI{0.2}{dex}$) for \mthree\ (GAEA, respectively). In the probed mass range, the two criteria give consistent \ac{sfr} thresholds below $\lmass\lesssim 11.1$ (10.7) for TNG300 (TNG100, respectively), but the \ac{ms}-based threshold deviates to lower values at higher masses.

The broadly similar threshold defined by both criteria in GAEA, \mthree\ and in TNG300 at masses where the main sequence can be robustly estimated suggests that the \cite{Franx_2008} criterion is appropriate to classify star-forming and quiescent galaxies in the simulated samples considered here. The discrepancy between the two criteria in TNG100 can be explained by a combination of small sample size and a high fraction of quenched galaxies at high masses, as discussed above. In the following we therefore denote galaxies as quiescent if they fulfil the \cite{Franx_2008} criterion.

\cite{Leja_2019_UVJ-sSFR} investigate the correlation between the location of galaxies at $0.5<z<2.5$ in a UVJ restframe color plane and their \ac{ssfr} and calibrate the separation in the UVJ plane for different thresholds of \ac{ssfr}. According to this calibration, the UVJ criterion from \cite{Williams_2009} roughly corresponds to a \ac{ssfr} threshold of $-9.5< \lssfr < -10.0 $, similar at face value to the \ac{ssfr} threshold adopted here.

\subsection{Quiescent fractions in simulations}
\label{sec:passivefractions}
In Figure~\ref{fig:mainsequence} we show quiescent fractions at $\lmass>11$, adopting the \cite{Franx_2008} criterion for quiescence as discussed above, as a function of stellar mass for different apertures, and compare with observational results at $2.5 < z< 3.0$ from \cite{Ilbert_2013}, \cite{Muzzin_2013_SMF}, \cite{Martis_2016_passive_fractions}, \cite{Davidzon_2017_SMF_COSMOS2015} and \cite{Sherman_2020_passive_fractions}. Uncertainties are obtained by calculating the binomial confidence intervals following \cite{Cameron_2011_binomial_confidence_interval}. At $\lmass>11$ we find a quiescent fraction of $\SI{27}{percent}$ for GAEA (with negligible statistical uncertainties given the large simulated volume) and ${19}\pm{3}\,\mathrm{percent}$ for \mthree\ with no significant dependence on the adopted aperture (among the choices discussed in Section~\ref{sec:sampleselection}).
Quiescent fractions for the same mass range in TNG are higher with ${44}\pm{2}\,\mathrm{percent}$ in TNG300 and ${34}\pm{5}\,\mathrm{percent}$ in TNG100, and more sensitive to the chosen aperture with larger apertures leading to smaller quiescent fractions. In TNG300 the difference between estimates with the smallest ($2\times r_{50}$) vs. the largest (including all particles) apertures is constant at $\approx 20$ percentage points over the full mass range, indicating that star formation is stronger in the galaxy outskirts \citep[see also][]{Donnari_2019, Merlin_2019_quiescent_galaxies_at_dawn_of_universe} and that quenching proceeds inside-out in TNG (\citealt{Nelson_2019_TNG50_outflows_AGN_supernovae}, \citealt{Nelson_E_2021_SFR_obs_vs_sim_offset}). This inside-out quenching is consistent with observations of 3D-HST galaxies at $z\approx1$ \citep[][]{Nelson_E_2021_SFR_obs_vs_sim_offset}. TNG300 quiescent fractions strongly increase with mass also in the $\lmass>11$ mass range investigated here, reaching values of $50-\SI{70}{percent}$ at $\lmass\approx11.5$. In TNG100 quiescent fractions decrease with increasing stellar stellar mass except for the smallest aperture ($2\times r_{50}^{\textrm{mass}}$). The different quiescent fractions in TNG300 and TNG100 are a result of resolution effects and sample variance \citep[see][]{Donnari_2021_quenched_fractions_AGN_environment_preprocessing, Donnari_2021_comparison_observations}.

Observed quiescent fractions from \cite{Ilbert_2013}, \cite{Muzzin_2013_SMF}, \cite{Martis_2016_passive_fractions} and \cite{Sherman_2020_passive_fractions} are close to those estimated for \mthree\ and GAEA and are systematically lower than in TNG300. Although at face value the agreement with TNG100 is better than with TNG300, because of the small sample size we cannot make a clear statement on this agreement. The quiescent fraction estimated in \cite{Davidzon_2017_SMF_COSMOS2015} is significantly lower than other observations\footnote{\cite{Davidzon_2017_SMF_COSMOS2015} explain such lower quiescent fractions as likely due to their use of the NUV$-r-J$ classification method \citep{Ilbert_2013} rather than UVJ, resulting in different sSFR thresholds. To investigate this we consider the COSMOS2015 \citep{LaigleCOSMOS2016} catalog on which \cite{Davidzon_2017_SMF_COSMOS2015} study is based. While \cite{Davidzon_2017_SMF_COSMOS2015} re-estimate photometric redshifts and stellar masses, we take the original masses and NUV-$r-J$ classification from \cite{LaigleCOSMOS2016} to compute quiescent fractions. These fractions are in good agreement with those from \cite{Muzzin_2013_SMF}, \cite{Martis_2016_passive_fractions} and \cite{Sherman_2020_passive_fractions}, so we conclude that the lower quiescent fractions in \cite{Davidzon_2017_SMF_COSMOS2015} might in fact also depend on the different redshift and stellar mass estimates.}.
Based on \ac{sed} fitting on optical (DES) and NIR (NEWFIRM, VISTA, CFHT, IRAC) data, \cite{Sherman_2020_passive_fractions} estimate quiescent fractions with three different criteria to select quiescent galaxies: UVJ color classification, $\lssfr<-11$ and $\mathrm{SFR}>\SI{1}{dex}$ below the \ac{ms}. They find consistent results with all criteria,
which also agree with quiescent fractions from \cite{Muzzin_2013_SMF}, \cite{Tomczak_2016_main_sequence_since_z4} and \cite{Martis_2016_passive_fractions} and in \mthree\ (see Figure~\ref{fig:mainsequence}). However, in contrast to their result, applying a quiescence threshold of $\lssfr=-11$ in the simulations (corresponding to $\approx\SI{2}{dex}$ below the \ac{ms}) results in quiescent fractions of $5-\SI{15}{percent}$ (depending on the adopted aperture) in \mthree, and $2-\SI{45}{percent}$ (strongly depending on aperture) in both TNG boxes (see Figure~\ref{fig:appendix_passive_fractions}). The quiescent fraction in GAEA decreases by only 1 percentage point because of the clear bimodality in the \ac{sfr} distribution.

In the following for the sake of brevity and readability we will present by default only results obtained with $\SI{30}{kpc}$ apertures around the center of mass, and explicitly comment as needed on results that depend on this choice.

\section{Stellar mass functions and number densities}
\label{sec:SMF}
\begin{figure*}[h!]
	\includegraphics[width=\textwidth]{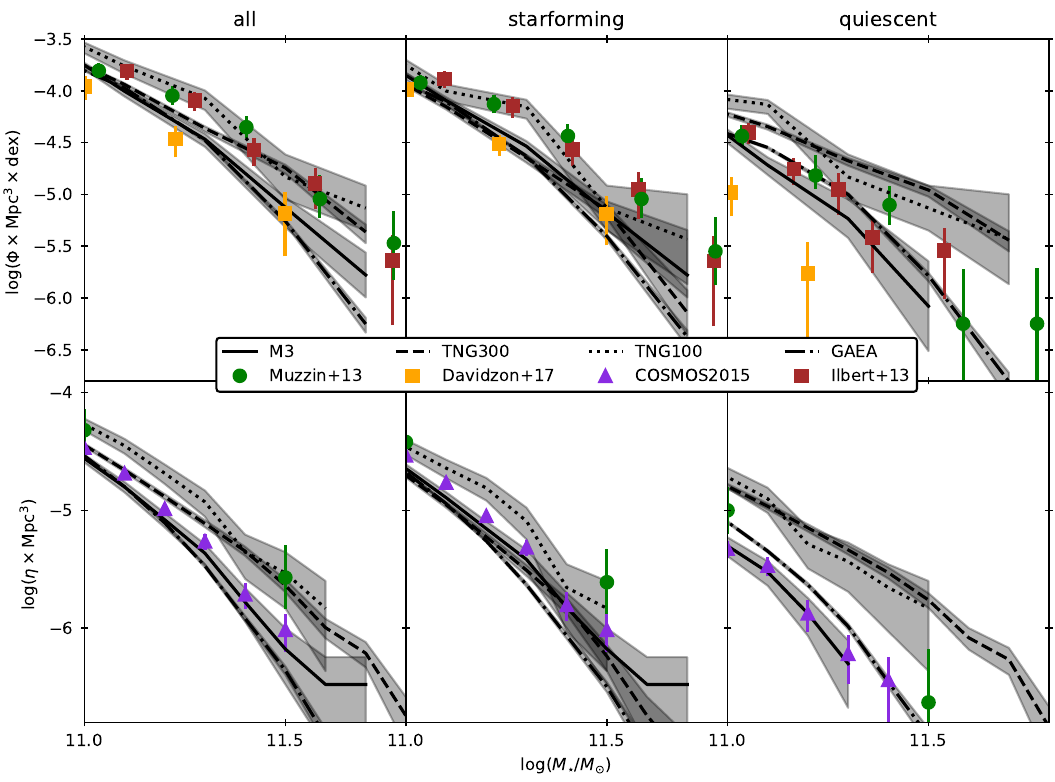}
	 \caption{Stellar mass functions and cumulative number densities at $z\approx 2.7$ in the studied simulations. {\it Top panels}: The \ac{smf} of all (left), star-forming (middle) and quiescent (right) galaxies from \mthree, TNG300 TNG100 and GAEA. Observational estimates are shown for comparison, as indicated. 
	 {\it Bottom panels:} The corresponding cumulative number densities of all, star-forming and quiescent galaxies compared with observational estimates as indicated. In all panels the grey shaded areas show poisson uncertainties.
	 Total \acp{smf} and number densities in the simulations are in overall good agreement with observations. Higher \acp{smf} and number densities for quiescent galaxies in TNG simulations with respect to \mthree, GAEA and observations reflect the higher quiescent fraction in TNG.
	 See Section~\ref{sec:SMF} for full details.}
	 \label{fig:massfunctions}
\end{figure*}

In the upper panel of Figure~\ref{fig:massfunctions} we compare observed \acp{smf}
at $2.5<z<3.0$ from \cite{Ilbert_2013}, \cite{Muzzin_2013_SMF} and \cite{Davidzon_2017_SMF_COSMOS2015} for all, quiescent and star-forming galaxies with those that we obtain for the simulations at $z\approx2.73$.
Given the stellar mass completeness limit of the observations considered here (Section~\ref{sec:observational_studies}) we focus on stellar masses larger than $10^{11}\,\Msun$.
Uncertainties are obtained by bootstrapping.
In the considered mass range the total \acp{smf} in \mthree\ and GAEA are in very good agreement and also with TNG300 up to $\lmass\approx11.3$. The stellar mass function in TNG100 is $\approx\SI{0.4}{dex}$ higher and overlaps with TNG300 at the high mass end.

The \ac{smf} of quiescent galaxies at $\lmass>11$ is significantly lower in GAEA and \mthree\ than in both TNG boxes (reflecting the high quiescent fraction in TNG already discussed in Section~\ref{sec:passivefractions}).
The best agreement between all simulations is seen in the \acp{smf} of star-forming galaxies where the lower fraction of star-forming galaxies in TNG is compensated by the slightly larger total \ac{smf}.

At $\lmass \gtrsim 11$ the observed \acp{smf} for all and star-forming galaxies from \cite{Ilbert_2013} and \cite{Muzzin_2013_SMF} are statistically consistent with TNG100 and $\approx\SI{0.4}{dex}$ higher than in \mthree\ and GAEA.
The total \ac{smf} from \cite{Davidzon_2017_SMF_COSMOS2015} is lower by $\approx\SI{0.4}{dex}$ compared to TNG100, \cite{Ilbert_2013} and \cite{Muzzin_2013_SMF} but in close agreement with GAEA and \mthree. The star-forming \ac{smf} from \cite{Davidzon_2017_SMF_COSMOS2015} matches GAEA, TNG300 and \mthree\ (because of their high fraction of star forming galaxies compensating the lower total \ac{smf}). 

The strongest differences in Figure~\ref{fig:massfunctions} can be seen for the \acp{smf} for quiescent galaxies. Results from \cite{Ilbert_2013} and \cite{Muzzin_2013_SMF} lie between (and broadly consistent with) those from the TNG boxes and \mthree, and show the best agreement with GAEA, while the \ac{smf} from \cite{Davidzon_2017_SMF_COSMOS2015} is much lower than in all simulations (reflecting their low quiescent fraction, as discussed in Section~\ref{sec:passivefractions}).

For quiescent galaxies with $\lmass\geq 11$ we find number densities of $\eta=(8.0\pm0.2)\times10^{-6}\,\mathrm{{Mpc}}^{-3},\ 
        (5.0\pm0.9)\times10^{-6}\,\mathrm{{Mpc}}^{-3},\ 
      (16.0\pm0.8)\times10^{-6}\,\mathrm{{Mpc}}^{-3}$ and $(19.2\pm3.8)\times10^{-6}\,\mathrm{{Mpc}}^{-3}$
for GAEA, \mthree, TNG300 and TNG100, respectively.
In the bottom panels of Figure~\ref{fig:massfunctions} we show cumulative number densities from all simulations and compare these results again with observational estimates from \cite{Muzzin_2013_SMF}. The agreement between the number densities reflects the agreement for the \acp{smf} shown in the upper panels. Because \cite{Davidzon_2017_SMF_COSMOS2015} do not estimate cumulative number densities, we estimate them from the COSMOS2015 \citep{LaigleCOSMOS2016} catalog (using the NUV-$r^+$ vs. $r^+-J$ quiescent vs. star-forming classification), finding good agreement with \mthree\ for all number densities (all, quiescent and star-forming galaxies), and relatively good agreement also with GAEA and TNG with the significant exception of the quiescent galaxy number density in TNG.

\begin{figure*}[h!]
	\includegraphics[width=\textwidth]{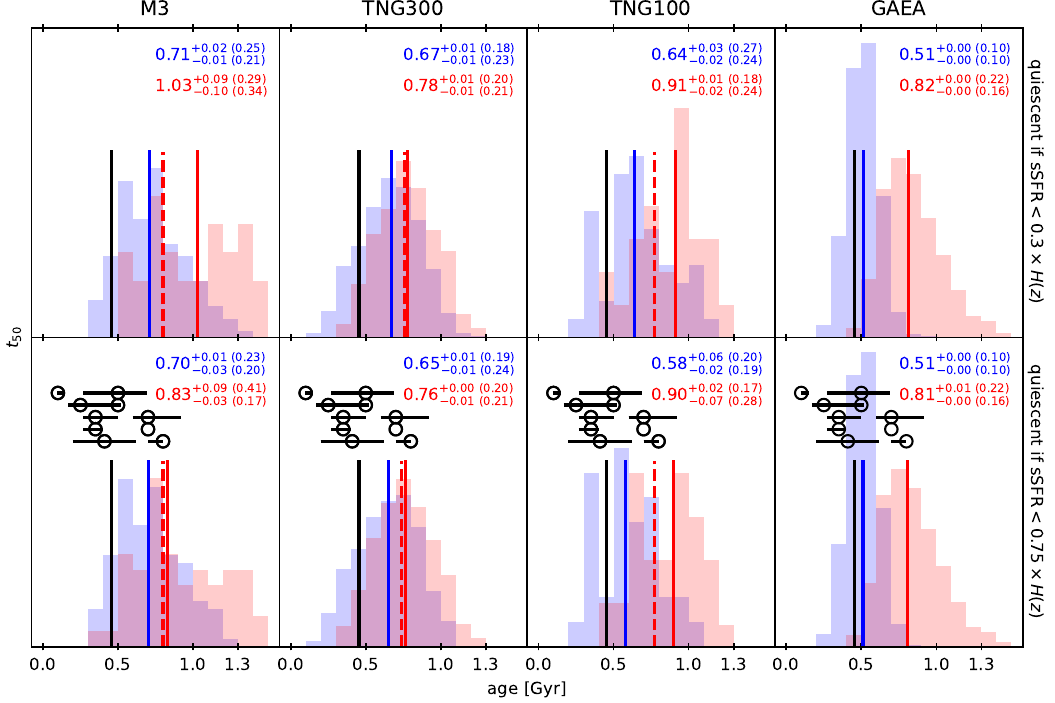}
	 \caption{Histograms of stellar ages ($t_{50}$) of quiescent (red) and star-forming (blue) galaxies in the studied simulations, as indicated. Median ages (shown by solid vertical lines) are given in each panel, together with their uncertainty and with the \ac{rms} of the age distribution. Dashed vertical lines show the median age of simulated quiescent galaxies with $H<22$ (see Section~\ref{sec:stellarages}). Black lines and circles show the median age and individual galaxy stellar ages for the observed sample of quiescent galaxies at $z\approx 2.7$ from \protect\cite{DEugenio_2020_paper}.
     Galaxies in the upper (lower) panels are classified as star-forming vs. quiescent based, respectively, on the \protect\cite{Franx_2008} criterion and shifting the \ac{ssfr} threshold up by a factor 2.5 (see Section~\ref{sec:stellarages}).
     }
	 \label{fig:age_distribution}
\end{figure*}

\begin{figure*}
	\includegraphics[width=\textwidth]{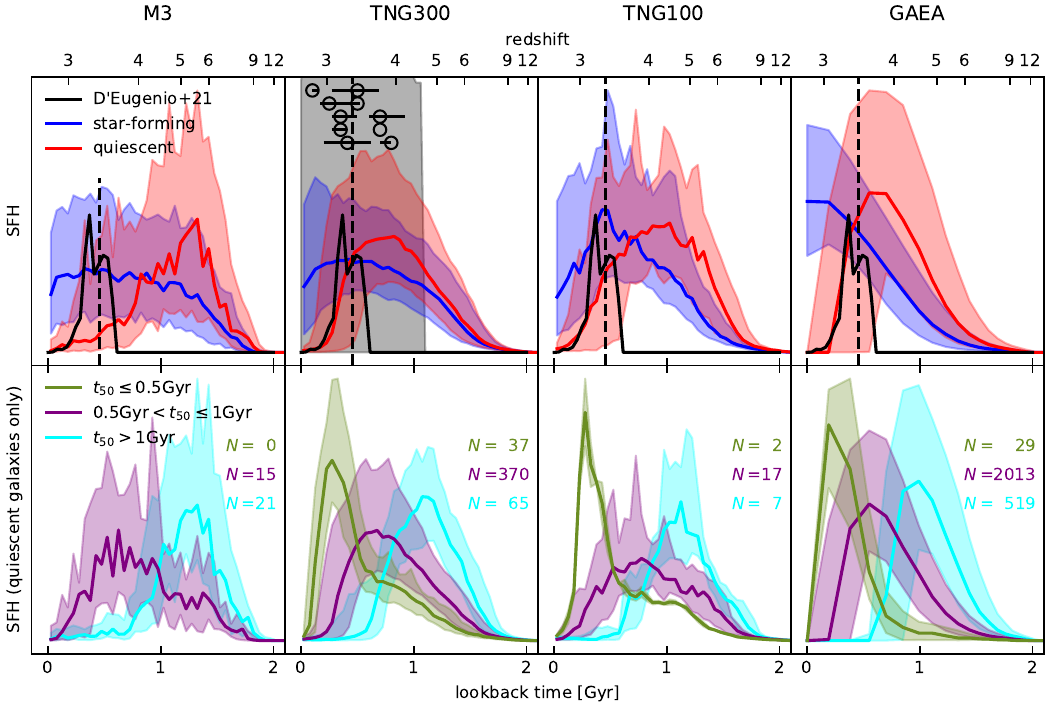}
	 \caption{\acp{sfh} shown as the fraction of the stellar mass at $z\approx 2.7$ formed per lookback-time interval in the studied simulations (as indicated), as a function of look-back time, for galaxies selected at $z\approx 2.7$. Top panels: average \ac{sfh} of all quiescent (red) and star-forming (blue) galaxies, classified according to their \ac{ssfr}. The black solid line shows the average \ac{sfh} of 10 observed quiescent galaxies from spectro-photometric modeling from \protect\cite{DEugenio_2020_paper}, and the black dashed line their median age ($t_{50}$). Black circles (top row, second panel) show individual ages for the same sample. Bottom panels: average \ac{sfh} of quiescent galaxies with $t_{50}\leq\SI{0.5}{Gyr}$ (green), $0.5< t_{50}\leq\SI{0.5}{Gyr}$ (violet) and $t_{50}>\SI{0.5}{Gyr}$ (cyan). In each panel, the number of galaxies in each age bin is reported in the corresponding color. Shaded areas around \acp{sfh} (including black line in top-middle panel) show the \ac{rms} of the corresponding distribution.
	 }
	 \label{fig:SFH}
\end{figure*}

\section{Stellar ages}
\label{sec:stellarages}
The age of the Universe at $z\approx 3$ is only $\approx\SI{2}{Gyr}$, but short star formation timescales at high redshift allow the existence of quiescent galaxies older than $\SI{1}{Gyr}$. However, spectroscopic studies confirming the most distant quiescent galaxies, at $3 \lesssim z<4$, find with very few exceptions only young galaxies (partly because of observational reasons, see below) with ages significantly below $\SI{1}{Gyr}$ \citep[][]{Gobat_2012, Schreiber_2018_NIR_spec_3z4, Saracco_2020_rapid_buildup_massive_ET_z3.4, Valentino_2020_Q_1.5years_after_binbang_and_progenitors, Forrest_2020_one_massive_galaxy_z3.5, Forrest_2020_survey_rapid_SF_and_quenching_early_universe, Kubo_2021_one_Q_galaxy_z3, Carnall_2022_JWST_Q_galaxies}.
For our observed sample at $2.4<z<3.2$, stellar ages are estimated by spectro-photometric modeling and are defined as the lookback time when half of the stellar mass of the galaxy was formed (hereafter, $t_{50}$). Based on such estimates, stellar ages in the observed quiescent sample are young, with a median $t_{50}$ of $\asuncunit{0.45}{0.10}{0.05}{Gyr}$ \citep[][]{DEugenio_2020_paper}.

To compare observed stellar ages of quiescent galaxies with those of galaxies in the simulations, we also calculate $t_{50}$ for the simulated galaxies (not accounting for mass losses, i.e. we consider the initial masses of the stellar particles). We note that, for both simulated and observed quiescent galaxies, the stellar ages as defined by $t_{50}$ purely refer to the formation time of the stars, regardless of their assembly history onto the given galaxy.
As previously discussed we consider in hydrodynamical simulations all particles within an aperture of $\SI{30}{kpc}$ at the snapshot redshift. Ages estimated in smaller apertures are slightly older but average ages are consistent within the uncertainties, and the aperture choice does not impact the results of the discussion. Histograms of $t_{50}$ for quiescent and star-forming galaxies together with the results for our observed quiescent sample are shown in Figure~\ref{fig:age_distribution}.
The average ages for quiescent galaxies in \mthree, TNG300 and TNG100 are $\asunc{1.03}{0.15}{0.11}, \asunc{0.78}{0.01}{0.01}$ and $\asuncunit{0.91}{0.10}{0.02}{Gyr}$ and for star-forming galaxies $\asunc{0.71}{0.01}{0.03}, \asunc{0.67}{0.01}{0.01}$ and $\asuncunit{0.64}{0.02}{0.04}{Gyr}$. For comparison the average mass-weighted age for quiescent galaxies in GAEA is $\SI{0.85}{Gyr}$, in good agreement with hydrodynamical simulations, while star-forming galaxies are slightly younger than in hydrodynamical simulations, with an average age of $\SI{0.49}{Gyr}$.
Although age distributions partly overlap, the median age of observed quiescent galaxies $t_{50}=\asuncunit{0.45}{0.10}{0.05}{Gyr}$ from \cite{DEugenio_2020_paper} is significantly younger than those of the simulated counterparts in all the considered simulations.

There may be several reasons for such age discrepancy on both the simulation and observation sides. We note first of all that age determinations from spectroscopic samples are expected to be biased towards younger ages, because the oldest galaxies are extremely difficult to observe even at very high masses with current instruments. Indeed, based on photometric observations older quiescent galaxies may actually exist \citep[e.g.,][]{Straatman_2014_quiescent_galaxy_population_z4, Carnall_2020_timing_quenching, Kalita_2021_old_masive_galaxy},
but while they may be elusive even in photometric studies, obtaining spectra to robustly confirm their nature and measure their ages is currently too expensive or unfeasible. Indeed, in \cite{Lustig_2021} and \cite{Deugenio_2020_letter} we have analysed our selection criteria for the observed sample and found a mild bias towards younger ages due to the applied H band selection for the spectroscopic follow-up. However, applying the same $H<\SI{22}{mag}$ cut to galaxies in the simulated sample has a marginal effect on the age distribution retrieved for simulated galaxies (see dashed lines in Figure~\ref{fig:age_distribution}), and thus we conclude that the expected age bias of spectroscopic samples is not likely to be a main explanation for the age discrepancy between observed and simulated quiescent galaxies.

Star formation histories in the simulations might be intrinsically different from those of real galaxies, resulting in too old ages with respect to observations. Furthermore, our specific analysis of the simulation might affect this result: we investigate the impact of our \ac{ssfr} criterion to select the quiescent sample on the age distribution of quiescent galaxies. The scatter of the \ac{ms} is approximately $\SI{0.3}{dex}$, independent of stellar mass at least up to $z\approx 3$ \citep[e.g.,][]{Whitaker_2012_main_sequence_z2.5, Speagle_2014_main_sequence, Tomczak_2016_main_sequence_since_z4, Pearson_2018_main_sequence_herschel}.
The \cite{Franx_2008} criterion discussed above thus identifies as quiescent only galaxies much below the \ac{ms} ($\gtrsim 3$ times the intrinsic scatter). This might limit our simulated quiescent samples to older ages. We therefore investigate this by reselecting quiescent galaxies including all sources with a \ac{sfr} already significantly below the \ac{ms} but that have not yet reached formal quiescence according to the \cite{Franx_2008} criterion, by relaxing the \ac{sfr} threshold for quiescence to $2\sigma$ ($\SI{0.6}{dex}$) below the \ac{ms}.
More specifically, following the discussion in Section~\ref{sec:sampleselection}, in particular with respect to the uncertainties in defining the \ac{ms} at high masses and the location of the \cite{Franx_2008} threshold $\approx10\times$ below the \ac{ms}, for the purpose of this check we define as quiescent those galaxies having a \ac{ssfr} of at most $2.5\times$ higher than the \cite{Franx_2008} threshold.
The average age of the selected quiescent population in the considered hydrodynamical simulations becomes younger with this relaxed cut, however, the change is marginal in both TNG boxes and about $~\SI{20}{percent}$ in \mthree, not strong enough to explain the discrepancy with the observed results. In GAEA this change has no impact on the average age because due to the pronounced bimodality in the \ac{sfr}-stellar mass distribution (see Figure~\ref{fig:mainsequence}) only $\SI{1}{percent}$ of the classifications are affected.
For a significant effect on the average age a much larger fraction of young star-forming galaxies would have to be classified as quiescent. In fact, Figure~\ref{fig:age_distribution} shows that the bulk of the star-forming population is anyway older than the average $t_{50}$ estimated in the observational sample by \cite{DEugenio_2020_paper}.

To investigate this discrepancy further we show in Figure~\ref{fig:SFH} the average \ac{sfh} of quiescent and star-forming galaxies in the studied simulations, that we obtain by averaging the fraction of formed mass in an interval of look-back time $\tlb$ for all galaxies. It can be seen that most of the star formation in the quiescent population happens at $\tlb>\SI{0.5}{Gyr}$, reflecting the relatively old ages retrieved in all simulations considered in this work. To estimate the $t_{50}$ of the observed galaxies from spectro-photometric modeling, \cite{DEugenio_2020_paper} marginalize over a set of constant, truncated constant, exponentially declining and delayed exponentially declining \acp{sfh}. The average onset of star formation estimated in \cite{DEugenio_2020_paper} occurs much later than in the simulations (only $\approx\SI{0.6}{Gyr}$ before observation epoch, see Figure~\ref{fig:SFH}).
To more specifically compare \acp{sfh} of quiescent galaxies we also show in Figure~\ref{fig:SFH} the average \ac{sfh} of simulated quiescent galaxies in different age bins.
Generally, in all simulations the maximum of star formation in quiescent galaxies is reached at larger lookback times than in star-forming galaxies.
In the TNG simulations about $\SI{8}{percent}$ of the galaxies have $t_{50}\leq \SI{0.5}{Gyr}$, $\SI{80}{percent}$ have $0.5<t_{50}\leq\SI{1.0}{Gyr}$, the remaining galaxies are older. On average their star-formation rate peaks $\approx 0.3$, $0.7$ and $\SI{1.1}{Gyr}$ before observation epoch, respectively. Except for galaxies in the oldest age bin, \SI{50}{percent} of the total stellar mass of the galaxies is already formed before the peak of star formation. Quiescent galaxies in \mthree\ are on average older (see Figure~\ref{fig:age_distribution}), with all quiescent galaxies being older than $\SI{0.5}{Gyr}$ and about $\SI{60}{percent}$ older than $\SI{1}{Gyr}$. In GAEA $\approx\SI{80}{percent}$ of the galaxies have ages between $\SI{0.5}{Gyr}$ and $\SI{1.0}{Gyr}$, $\approx\SI{20}{percent}$ are older than $\SI{1.0}{Gyr}$.

In this respect, we note that the $t_{50}$ estimated from spectro-photometric fitting, being estimated from galaxy light, is anyway preferentially biased towards younger stellar populations, and especially so for complex \acp{sfh} with a significant fraction of stellar mass formed at early times but with a significant recent burst, which are often not properly accounted for by the adopted \ac{sfh} libraries.
As a limiting case we therefore also calculate observed H band light-weighted ages for galaxies in hydrodynamical simulations ($\approx\SI{4300}{\angstrom}$ restframe). Although we stress that ages estimated as $t_{50}$ from spectro-photometric modeling \citep[e.g.,][]{DEugenio_2020_paper} formally aim at estimating mass-weighted ages, we consider here light-weighted ages as a limiting case where the impact of the recent \ac{sfh} on the age estimate is strongest.
The resulting ages are younger with a median of $\approx\SI{0.6}{Gyr}$ for quiescent and $\SI{0.3}{Gyr}$ for star-forming galaxies in all simulations.

To further probe the impact of the \ac{sed} modeling procedure on the retrieved age estimates of observed galaxies in comparison with what we obtain for simulated galaxies, we thus estimate ages for simulated galaxies using multi-band photometry in the COSMOS2015 photometric bands from synthetic, dust-reddened \acp{sed}, computed as described in Section~\ref{sec:luminosities}, and accounting for photometric uncertainties equivalent to those in the COSMOS2015 catalog. We adopt here for reference the \ac{sed} modeling carried out by \cite{DEugenio_2020_paper}, and apply the same procedure to the synthetic photometry of simulated galaxies.
We stress that the comparison of this modeling results with observational estimates cannot be interpreted in exact, quantitative terms, since the estimated ages depend on the details of the adopted procedure to derive \acp{sed} for the simulated galaxies. Although the comparison thus remains qualitative (and for this reason we do not show actual estimates in the figure), we nonetheless find that the $t_{50}$ estimates obtained with such \ac{sed} modeling do not match the estimated mass-weighted age, but are skewed to lie in between light-weighted and mass-weighted ages. This suggests that possible biases in the observational determination of stellar ages might at least partly contribute to the tension between young ages found for quiescent galaxies at $z\approx 3$ in current observational studies and older ages from numerical simulations.

To date only few quiescent galaxies at $z\approx 2.7$ are spectroscopically confirmed, so that our comparison is limited to the sample of 10 galaxies from \cite{DEugenio_2020_paper}. However, we note that although the typically young ages of (38, overall) observed quiescent galaxies at $3\lesssim z < 4$ from \cite{Gobat_2012},
\cite{Schreiber_2018_NIR_spec_3z4}, \cite{Saracco_2020_rapid_buildup_massive_ET_z3.4}, \cite{Valentino_2020_Q_1.5years_after_binbang_and_progenitors}, \cite{Forrest_2020_one_massive_galaxy_z3.5}, \cite{Forrest_2020_survey_rapid_SF_and_quenching_early_universe},
\cite{Carnall_2022_JWST_Q_galaxies} are in good agreement with ages found by \cite{DEugenio_2020_paper}, $\SI{90}{percent}$ of these galaxies are at $z>3$, resulting in a median $z\approx 3.4$. Assuming pure passive evolution once these galaxies have quenched, the average age of their descendants at $z\approx 2.7$ is in good agreement with average ages found in simulations. Although this suggest that there may be galaxies that quench at sufficiently early times to be consistent with stellar ages of simulated galaxies at $z\approx 2.7$, it remains unclear how this comparison is affected by selection bias, and how representative these galaxies are for the overall population of quiescent galaxies at $z\approx 2.7$.

Therefore, although at face value there is a discrepancy in the ages of quiescent galaxies in observed samples and in simulations, a quantitative estimate of the actual significance of this discrepancy is limited by potential observational biases (in sample selection as well as in the estimation of stellar ages) as well as in possible inconsistencies in the analysis of the simulated galaxies with respect to the observed ones (including star formation histories and dust attenuation prescriptions).

\section{UVJ selection of high-redshift quiescent galaxies}
\label{sec:uvj}
In \cite{Lustig_2021} we analysed the UVJ restframe color plot of massive ($\lmass>11.1$) galaxies at $2.5 < z < 3.0$ from the \cite{Muzzin_2013} and COSMOS2015 \citep{LaigleCOSMOS2016} catalogs. The combination of photometric uncertainties across the UVJ diagram and the distribution of $\SI{24}{\micro\metre}$-detected sources suggested a potentially significant contamination of UVJ-quiescent samples by dusty star-forming sources. According to our analysis potential contaminants amount to $\approx \SI{20}{percent}$ of the $\lmass>11.1$ UVJ quiescent galaxy sample at this redshift, preferentially affecting the UVJ region typically populated by older quiescent galaxies \citep[for full details see][]{Lustig_2021}.
In this section we investigate the purity and completeness of a UVJ selected quiescent galaxy sample as identified in the simulations mimicking the selection criteria and photometric uncertainties as in the observed case. We note already here that in the attempt to reproduce the observed selection we need to rely on assumptions concerning dust attenuation of star-forming and quiescent populations (see Section~\ref{sec:luminosities}), and we will neglect the impact of \acp{agn} on the galaxy \acp{sed}. We also note that by construction the following results apply for photometric uncertainties typical of the COSMOS2015 catalog.

\begin{figure*}
	\includegraphics[width=\textwidth]{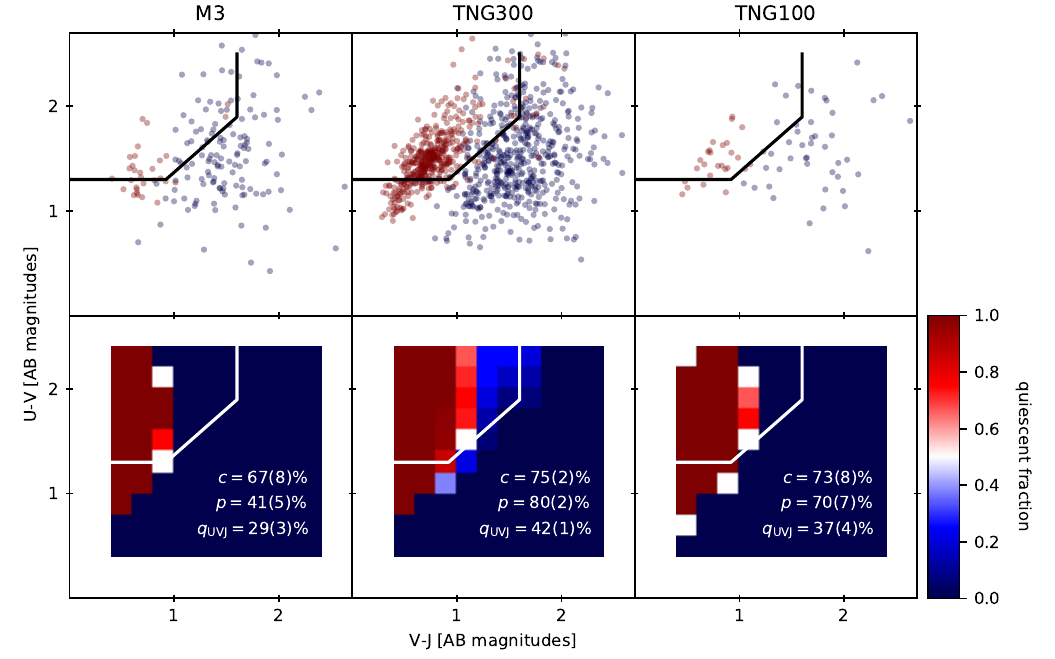}
	 \caption{The restframe $U-V$ vs. $V-J$ color diagram for simulated galaxies at $z\approx 2.7$, adopting prescriptions for dust attenuation and photometric uncertainties corresponding to the COSMOS2015 \protect\citep{LaigleCOSMOS2016} catalog. See Section~\ref{sec:luminosities} for full details.
	 Top panels: a random realization of the diagram for the studied simulations, as indicated, including scatter in dust attenuation and photometric uncertainties as detailed in Section~\ref{sec:luminosities}. Quiescent and star-forming galaxies according to the \protect\cite{Franx_2008} criterion are shown with red and blue dots, respectively. Bottom panels: the average fraction of quiescent galaxies (according to the \protect\cite{Franx_2008} criterion) over 1000 realizations, as a function of the position across the diagram, as indicated by the color bar.
	 In each of the bottom panels, the corresponding estimated completeness ($c$) and purity ($p$) of the UVJ-selected quiescent sample are given, together with the UVJ-derived quiescent fraction $q_{\mathrm{UVJ}}$ (error bars give the \ac{rms} across the different realizations).}
	 \label{fig:uvj}
\end{figure*}
 
We calculate 1000 realisations of UVJ restframe colors for all simulated galaxies in the studied hydrodynamical simulations. For each realization we apply dust attenuation by randomly assigning an $\Av$ to each galaxy according to its classification based on the \cite{Franx_2008} \ac{ssfr} criterion, as explained in Section~\ref{sec:luminosities}, accounting for the scatter in $\Av$ for both star-forming and quiescent populations. The photometry of each galaxy is perturbed to match the photometric uncertainties of the COSMOS2015 catalog (see Section~\ref{sec:luminosities}). We then divide the UVJ plane into bins of $V-J$ and $U-V$ and calculate for each bin and realization the fraction of quiescent galaxies according to the \cite{Franx_2008} criterion.
One random realization of the UVJ diagram for all considered hydrodynamical simulations is shown in Figure~\ref{fig:uvj} (top panels). Figure~\ref{fig:uvj} (bottom) also shows the average quiescent galaxy fraction as a function of the location in the UVJ diagram.
We also show the purity and completeness of a UVJ-selected \citep{Williams_2009} quiescent galaxy sample with respect to the \ac{ssfr} selection \citep{Franx_2008}, as well as the overall UVJ quiescent galaxy fraction.
If considering the impact of the different dust attenuation prescriptions considered in Section~\ref{sec:luminosities} we find absolute differences for purity, completeness and overall UVJ quiescent fractions of at most 5, 4, and 3 percentage points, respectively.

In the lower panel in Figure~\ref{fig:mainsequence} we show quiescent fractions in the simulations as derived from UVJ classification with the considered photometric uncertainties and assumptions on dust attenuation. For an aperture of $\SI{30}{kpc}$ (see Section~\ref{sec:sampleselection}) the overall UVJ quiescent fraction of $\lmass>11$ galaxies in \mthree\ is $\approx \SI{30}{percent}$ (about $10$ percentage points higher than according to the \ac{ssfr} classification) and $\SI{40}{percent}$ in the TNG simulations (consistent with \ac{ssfr} classification). Quiescent fractions in TNG simulations strongly depend on the considered aperture and are between $\approx\SI{35}{percent}$ (considering all bound particles) and $\approx\SI{55}{percent}$ (considering particles within $2\times r_{50}$, see Section~\ref{sec:sampleselection}). Quiescent fractions for different apertures in \mthree\ differ by at most 4 percentage points.

However, there are significant differences between the \ac{ssfr}- and UVJ-selected quiescent samples, that reflect in the completeness and purity of the UVJ-selected sample by comparison to the \ac{ssfr}-based quiescence definition. The quiescent part of the UVJ diagram shows strong contamination from star-forming galaxies at $V-J\gtrsim 1$ in all simulations. In \mthree\ the overall purity with respect to the \cite{Franx_2008} \ac{ssfr} criterion of a UVJ selected quiescent galaxy sample with $\lmass>11$ is only $\approx \SI{41}{percent}$ and the completeness (with respect to the full \ac{ssfr}-selected quiescent sample) is $\approx\SI{67}{percent}$ (see further discussion below).
As expected due to the intrinsic higher quiescent fraction (see Section~\ref{sec:passivefractions}), the contamination in the TNG simulations is lower with a purity of the overall UVJ-quiescent sample of $\approx\SI{80}{percent}$ in TNG300 and $\SI{70}{percent}$ in TNG100. The completeness is $\approx\SI{75}{percent}$ in both TNG boxes. We stress again that this only reflects the higher intrinsic quiescent fraction in TNG with respect to \mthree.
Considering the higher intrinsic quiescent fraction in TNG simulations we find that the actual difference in purity and completeness with respect to \mthree\ is only $\approx10$ percentage points.

Following our discussion in Section~\ref{sec:stellarages} we also repeat the calculation applying the relaxed \ac{ssfr} cut (\ac{sfr} 4 times below the \ac{ms}) for the classification of quiescent galaxies. The corresponding UVJ color plots are shown in Figure~\ref{fig:appendix_uvj_relaxed_ssfr_cut}. While the UVJ-derived quiescent fraction does not significantly change (within $4$ percentage points with no systematic shifts), the purity of the UVJ quiescent sample increases by $10-16$ percentage points, reaching $\SI{57}{percent}$ in \mthree\ and $82-\SI{90}{percent}$ in TNG, because of an overall increase of the quiescent population due to young quenched sources with \ac{ssfr} between 4 and 10 times below the \ac{ms} (this result is stable against the adopted dust attenuation prescription from Section~\ref{sec:luminosities}).
This might suggest that a non-negligible fraction of the observed UVJ-quiescent population \citep[within the typically adopted UVJ-quiescent region, e.g.,][]{Williams_2009} is made of intrinsically very young sources spread throughout the UVJ quiescent region by dust attenuation. 

From our analysis of the UVJ diagram of $\lmass>11.1$ galaxies at $2.5<z<3$ in \cite{Lustig_2021}, considering photometric uncertainties and $\SI{24}{\micro\metre}$ detections, we estimated the purity of a UVJ selected quiescent sample to be in the order of $\SI{80}{percent}$ with an increasing contamination towards the region typically populated by older quiescent galaxies, the latter in qualitative agreement with our findings from simulations.

As discussed above, our modelling of the UVJ diagram for simulated galaxies with both considered \ac{ssfr} thresholds would suggest a higher purity of the UVJ-quiescent sample for TNG, in better agreement with purity estimates derived from observations, with respect to \mthree. This higher purity results from the substantially higher intrinsic quiescent fraction in TNG with respect to \mthree\ (see Figure~\ref{fig:mainsequence}).
In this respect we note that in TNG simulations the "observationally estimated" UVJ-derived quiescent fractions at $\lmass>11$ are $\approx\SI{40}{percent}$ and differ by at most $3$ percentage points from intrinsic (\ac{ssfr}-based) quiescent fractions, while in \mthree\ UVJ-derived quiescent fractions are $\approx\SI{30}{percent}$ ($\approx10$ percentage points higher than intrinsic quiescent fractions, see bottom panels of Figure~\ref{fig:mainsequence}). 

Independent of the exact \ac{ssfr} classification criterion and for all dust attenuation parametrizations that we considered (see Section~\ref{sec:luminosities}), the UVJ classification criterion in its standard form adopted at lower redshifts \citep[e.g.,][]{Williams_2009} might not be ideal for a high completeness and purity sample of massive high redshift galaxies \citep[with the photometric uncertainties considered here, see also e.g., ][]{Merlin_2018}.
Investigating a sample of spectroscopically confirmed quiescent galaxies at $3<z<4$, \cite{Schreiber_2018_NIR_spec_3z4} find that an increasing fraction of massive quiescent galaxies at high redshift with a \ac{sfr} reduced by at least $\SI{90}{percent}$ with respect to their main formation phase has not yet entered the quiescent part of the standard UVJ selection. For more complete samples of massive quiescent galaxies at high redshift they suggest to adjust the selection criteria by removing the $U-V > 1.3$ constraint that is used to avoid contamination with star-forming galaxies but is less relevant for massive samples where photometric uncertainties are smaller and star-forming sources are typically more dusty. Also \cite{Forrest_2020_survey_rapid_SF_and_quenching_early_universe} find that a standard UVJ diagram does not provide a pure or complete selection of quiescent galaxies for massive samples at high redshift. Figure~\ref{fig:uvj} shows the UVJ diagram for all three simulations according to the model described above. Our findings support the suggestion of removing the $U-V>1.3$ constraint. Furthermore, with the typical depth of the COSMOS2015 catalog our modeling (and we stress again the caveats deriving from making assumptions on dust attenuation, neglecting AGN contribution, and relying on \acp{sfh} from the considered simulations) suggests that strong contamination from dusty star-forming sources in the UVJ quiescent region at $V-J\gtrsim 1$ may reduce significantly the purity of such a photometrically selected sample, potentially biasing derived properties for the quiescent population.
If in our analysis we remove the $U-V>1.3$ constraint and we only consider the quiescent population at $V-J<1$, the purity of the selected samples increase
from $\lesssim 50$ to $80-\SI{90}{percent}$ while completeness remains at the $\approx\SI{90}{percent}$ level.

\section{Morphologies}
\label{sec:morphology}
In \cite{Lustig_2021} we have analysed the morphologies of quiescent galaxies in our observed sample by fitting \cite{Sersic_1963, Sersic_1968} profiles to the \acl{wfc3} F160W band images of our targets. We found compact structures with a median effective radius along the major axis of $\re=\asuncunit{1.4}{0.2}{0.9}{kpc}$, consistent with previous work suggesting size evolution by nearly an order of magnitude for massive quiescent galaxies across the redshift range $0<z<3$. We found high \sersic\ indices with an average of $n=\asunc{4.5}{1.4}{0.3}$, suggesting that the correlation between early-type morphology and quiescence already occurs at high redshift, with massive quiescent galaxies being typically bulge dominated at $z\approx3$.
In this section we investigate whether structural properties of simulated galaxies are correlated with quiescence, and thus morphological differences between star-forming and quiescent galaxy populations at $z\approx 2.7$ are predicted in simulations. Because the internal structure of galaxies in \acp{sam} is not resolved, we focus here on the comparison with hydrodynamical simulations. An analysis of morphological properties of galaxies in GAEA at low redshift and their evolution at higher redshift can be found in \citet[][]{Zoldan_2018_sizes_angular_momenta_GAEA, Zoldan_2019_evolution_angular_momentum_sizes_GAEA}.

Because of the different inherent properties and statistical and systematic uncertainties of the probe of galaxy structure in observations and simulations, we cannot analyse the morphologies of simulated galaxies exactly as we do with actual observations.
Indeed the strongest constraints on the profiles of observed galaxies come from the inner part of the surface brightness profile with the highest \ac{snr}, while the outskirts are progressively more and more dominated by noise.
In simulations the gravitational softening length modifies particle-particle interactions at small scales to avoid too close encounters of particles. This smooths out the distribution of particles in the central part of simulated galaxies where the density is very high and, from a modeling perspective, is not analogous to the smoothing by the \ac{psf} occurring in actual images. When analysing galaxies much larger than the softening length the central part can be excluded to fit \sersic\ profiles \citep[e.g.,][]{Remus_2021_accretion}. However, at higher redshift, where galaxies are at fixed mass much smaller than in the local Universe, this kind of fit may be very sensitive to the size of the masked part.
Following a range of tests to explore the impact of this effect, we therefore decided to use non-parametric descriptions of the morphologies that we describe in the following. We stress here that results discussed in the following are thus, by construction, not based on the direct, quantitative comparison of similarly estimated properties on observed and simulated galaxies. We rather attempt to investigate with a range of probes whether we can find structural differences between star-forming and quiescent populations in the simulated samples at this redshift. In this perspective, although we do not have observational studies of the angular momenta of statistical samples of galaxies at this redshift today, we investigate also the angular momenta of star-forming and quiescent galaxies in the simulations, as a potential tracer of different structural properties.

\subsubsection{Axis ratios}
In a first step we iteratively calculate axis ratios for randomly projected simulated galaxies following the equations in \citet[][]{sextractor} modified for our purpose where the positions and emitted light of individual particles are known. Briefly, in each step we calculate the second moments of the projected particle positions, weighted by their simulated observed H band emission:

 \begin{equation}
\overline{x_j x_k} = \frac{\sum_{i} m_i x_{i, j} x_{i, k}}{\sum_{i} m_i} - \left(\frac{\sum_{i} m_i x_{i, j}}{\sum_{i} m_i}\right)
\left(\frac{\sum_{i} m_i x_{i, k}}{\sum_{i} m_i}\right),
 \end{equation}
 
 where $x_{i,j}$ is the $j$-th component of the position of particle $i$ and $m_i$ the weight. The semimajor ($A_+$) and semiminor axes ($A_-$) can then be calculated as:

\begin{equation}
 {A_{\pm}}^2  = \frac{\overline{x_1^2}+\overline{x_2^2}}{2}
     \pm \sqrt{\left(\frac{\overline{x_1^2}-\overline{x_2^2}}{2}\right)^2 + \overline{x_1x_2}^2}.
 \end{equation}

The position angle $\theta$ is given by the following equation:
\begin{equation}
\tan 2\theta = 2 \frac{\overline{xy}}{\overline{x^2} - \overline{y^2}}.
\end{equation}

We then define for a particle $i$ its distance $\rellipi$ from the center of the galaxy accounting for the ellipticity as:

\begin{equation}\label{eq:ellipticity}
\rellipi=\left|\begin{pmatrix}
1 & 0\\
0 & q^{-1}
\end{pmatrix}
R(-\theta)(\mathbf{x}_i - \mathbf{x}_{\mathrm{c}})\right|
\end{equation}
where $q=A_{-}/A_{+}$ is the axis ratio of the galaxy, $\mathbf{x}_\mathrm{c}$ its center, $\mathbf{x}_i$ the position of particle $i$ and the matrix $R(-\theta)$ rotates the positions by $-\theta$. 

Given the typical depth of the images we used for the modeling of observed galaxies, the surface brightness profiles of the observed targets is equal to the background \ac{rms} at a distance of on average $\SI{6}{kpc}$ from the center. For this reason, to more closely probe the axis ratios estimated for the observed galaxies, we exclude all particles with $\rellipi>\SI{6}{kpc}$ and repeat the calculation of the axis ratio and the rotation angle until convergence (considering only particles with $\rellipi<\SI{3}{kpc}$ rather than $\SI{6}{kpc}$ increases the axis ratios only marginally by $0.02$ on average).

\subsubsection{Sizes and Concentration}
The effective radius of a \sersic\ profile contains $
\SI{50}{percent}$ of the total light of the galaxy. To compare with the measured sizes for our observed quiescent sample and with results from \cite{VanDerWel2014} on a larger statistical galaxy sample though with limited statistics for massive quiescent galaxies at this redshift, we measure from the simulated data half-light radii of the semimajor axis within an elliptical $\SI{30}{kpc}$ aperture for the simulated galaxies using the definition of the radius accounting for ellipticities from equation~\ref{eq:ellipticity}. 
As a measure of the concentration we use the definition:
\begin{equation}
C=5\times\log(r_{\mathrm{o}} / r_{\mathrm{i}})
\end{equation}
from \cite{Kent_1985_concentration}. In most works the outer ($r_{\mathrm{o}}$) and inner ($r_{\mathrm{i}}$) radii are defined as $r_{80}$ and $r_{20}$ (containing 80 and $\SI{20}{percent}$ of the total light, respectively).
Since for about half of the galaxies in the TNG300 simulation at $z=2.73$ the estimated $r_{20}$ are smaller than the softening length these radii may be biased by the softening and we therefore use instead as inner radius the half-light radius, which is larger than the softening length for more than $\SI{95}{percent}$ of the relevant sample. We include in the further analysis also the $\SI{5}{percent}$ galaxies with a half-light radius smaller than the softening length, removing them has no significant impact on estimated average properties. Both radii are measured along the semi-major axis.
All galaxy sizes defined above clearly depend on the aperture chosen as representative of the total galaxy flux (or mass). If measuring $r_{50}$ ($r_{80}$) within an aperture of 2 times the half-light radius rather than in $\SI{30}{kpc}$ apertures as discussed above, sizes of individual galaxies would be on average smaller by $\SI{35}{percent}$ ($\SI{60}{percent}$). The concentration of individual galaxies would decrease by $\approx 1$ on average, however, the significance of the difference of average concentrations between star-forming and quiescent galaxies is not affected.

For a closer comparison between observations and simulations, we compute concentrations corresponding to the $r_{80}/r_{50}$ ratio also for our observed quiescent sample, based on \sersic\ profiles from \cite{Lustig_2021}. 
If only the central $\SI{30}{kpc}$ of the \sersic\ profiles are considered rather than the full profile, concentrations are lower by $<0.01$ for 6 galaxies out of 10. For 4 galaxies with \sersic\ index $\gtrsim 5$ the concentration decreases by $0.24-0.86$.

\subsubsection{Specific angular momentum and $b$-value}

We calculate the 2-dimensional specific angular momentum within elliptical apertures for all simulated galaxies as:
\begin{equation}
j = \left| \frac{\sum_i m_i\, \mathbf{x}_i \times \mathbf{v}_i}{\sum_i m_i}\right|
\end{equation}
where $m_i$ is the emitted light of particle $i$ in the observed H band, $\mathbf{x}_i$ its projected position with respect to the center of light and $\mathbf{v}_i$ its velocity with respect to the global light-weighted velocity of the galaxy. Specific angular momenta estimated within $2\times r_{50}$ instead of $\SI{30}{kpc}$ are on average $\SI{45}{percent}$ smaller. However, the significance of the average angular momentum difference between star-forming and quiescent galaxy populations is not impacted by the choice of the aperture.

In addition we calculate for all galaxies the $b$-value, defined as the logarithmic specific angular momentum at a pivot stellar mass of $1\,\Msun$, assuming $j\propto \Mstar^{2/3}$ \citep[][]{Teklu_2015_magneticum_angular_momentum_galaxy_dynamics}: 
\begin{align}
    b &= \log(j \times\, \mathrm{s\,/\, km\,/\,kpc})-2/3\times\lmass\\
    & =\log\left(f_j/f_{\star}^{2/3}\right)+c,
\end{align}
where $f_{\star}$ is the baryon conversion efficiency, $f_j$ is the fractional $j$ net retention factor and $c$ is a constant \citep[see][]{Romanowsky_2012_AngularMomentum}.

\subsection{Structural properties in relation to quiescence}
\label{sec:morphology_discussion}
\begin{figure*}
	\includegraphics[width=\textwidth]{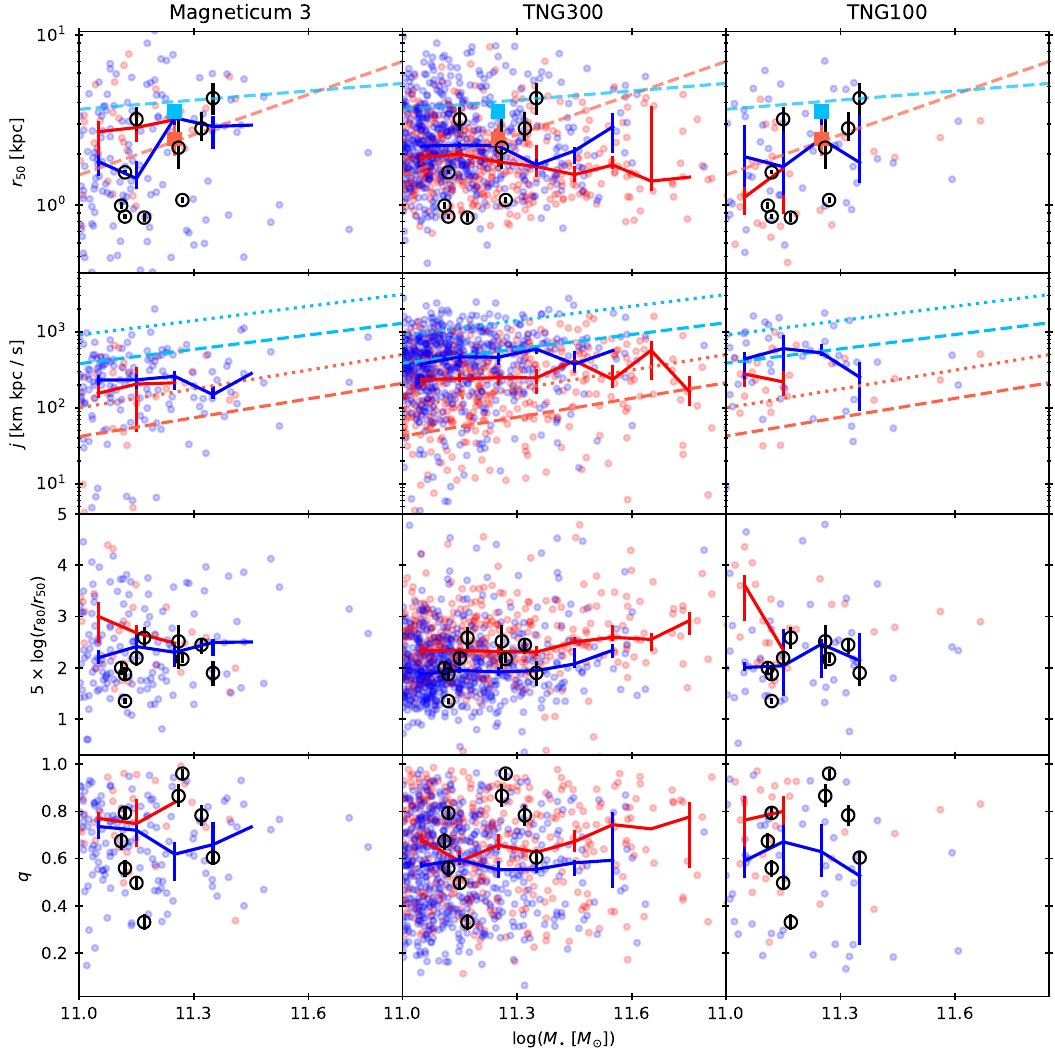}
	 \caption{Morphological properties (projected) of quiescent (red) and star-forming (blue) galaxies at $z\approx 2.7$ as a function of stellar mass in the studied simulation as indicated.
	 Individual galaxies are shown with filled circles, solid lines indicate median values estimated in $\SI{0.1}{dex}$ bins of stellar mass if at least 5 galaxies fall in the bin. Black open circles show observational results for quiescent galaxies from \protect\cite{Lustig_2021}. In the first-row panels we show half-light (observed H band) radii along the semi-major axis. The dashed red (blue) line and square show the best-fit mass-size relation and average size in the $11.0<\lmass<11.5$ bin for observed quiescent (star-forming) galaxies at $2.5<z<3.0$ from \protect\cite{VanDerWel2014}. In the second-row panels we show specific angular momenta. Light blue (light red) lines are references for quiescent (star-forming) galaxies from \protect\cite{Fall_2018_AngularMomentum} scaled to the redshift of the simulation by $j$ vs. $z$ relations from \protect\citet[][dashed lines]{Swinbank_2017_angular_momentum_evolution} and \protect\citet[][dotted lines]{Lagos_2017_angular_momentum_evolution_EAGLE}. Panels in the third and fourth rows, respectively, show galaxy concentration and axis ratio as a function of stellar mass.}
	 \label{fig:morphology}
\end{figure*}

\begin{figure*}
	\includegraphics[width=\textwidth]{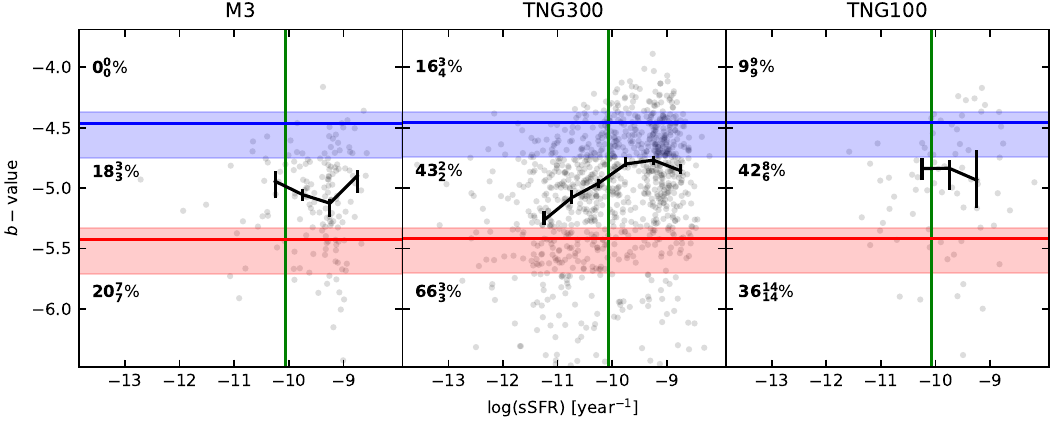}
	\caption{The $b$-value (specific angular momentum scaled to a pivot mass of $1\,\Msun$ assuming $j\propto M^{2/3}$) for all galaxies at $z\approx 2.7$ as a function of \ac{ssfr} in the studied simulations as indicated. Grey dots show individual galaxies. Blue (red) lines show the average $b$-value of observed nearby spiral (elliptical) galaxies from \protect\cite{Romanowsky_2012_AngularMomentum} scaled to $z\approx 2.7$ assuming $j\propto(1+z)^{-1/2}$ \protect\citep[][]{Obreschkow_2015_angular_momentum_scaling}. The lower (upper) limit of the shaded areas indicates the \bvalue\ if scaled to $z\approx 2.7$ adopting the relation from \protect\cite{Swinbank_2017_angular_momentum_evolution} \protect\citep[][]{Lagos_2017_angular_momentum_evolution_EAGLE}. Green vertical lines show the \ac{ssfr} threshold for quiescence adopting the \protect\cite{Franx_2008} criterion. The black solid lines show the median $b$-value in bins of \ac{ssfr}. Numbers indicate the quiescent fractions in the three ranges below the red solid line / above the blue solid line / in between.
	}
	\label{fig:bvalue}
\end{figure*}

The morphological parameters discussed above are shown as a function of stellar mass in Figure~\ref{fig:morphology}. The average morphological parameters for $\lmass>11$ quiescent and star-forming galaxies are listed in Table~\ref{tab:MorphologicalProperties}. In Figure~\ref{fig:bvalue} we show $b$-values as a function of \ac{ssfr}. Uncertainties on the average properties and on the difference between the two populations are obtained with bootstrapping. The median mass of quiescent and star-forming galaxies in the considered simulations differs by at most $\SI{0.06}{dex}$ and this difference does not impact the comparison.
Due to the small sample size in the TNG100 simulations the uncertainties on the average parameters are very large. Because of the consistency with TNG300 results we only consider the latter and \mthree\ in the following.

\newcommand{\morphtablecaption}{Average half-light radii ($r_{50}$), specific stellar angular momentum ($j$), concentration ($C_{58}$) and axis ratio ($q$) for the $\lmass>11$ samples of quiescent (Q) and star-forming (SF) galaxies and their ratio.}

\begin{table}
    \renewcommand{\arraystretch}{1.5}
	\centering
	\caption{\morphtablecaption}
	\label{tab:MorphologicalProperties}
	\begin{tabular}{ccrrr}
		\hline
              && M3 & TNG300 & TNG100\\ \hline    & Q & $\asunc{3.2}{0.4}{0.2}$ & $\asunc{1.8}{0.1}{0.1}$ & $\asunc{1.6}{0.5}{0.2}$ \\
    $r_{50}$& SF & $\asunc{1.9}{0.2}{0.2}$ & $\asunc{2.20}{0.04}{0.07}$ & $\asunc{2.2}{0.4}{0.2}$ \\
    \lbrack kpc\rbrack& ratio & $\asunc{1.7}{0.3}{0.2}$ & $\asunc{0.81}{0.03}{0.04}$ & $\asunc{0.7}{0.2}{0.2}$ \\
\hline
    & Q & $\asunc{172}{35}{64}$ & $\asunc{227}{20}{18}$ & $\asunc{272}{50}{84}$ \\
    $j$& SF & $\asunc{219}{23}{11}$ & $\asunc{411}{15}{32}$ & $\asunc{405}{26}{49}$ \\
    \lbrack km kpc / s\rbrack& ratio & $\asunc{0.8}{0.2}{0.3}$ & $\asunc{0.5}{0.1}{0.1}$ & $\asunc{0.7}{0.1}{0.3}$ \\
\hline
    & Q & $\asunc{2.6}{0.1}{0.2}$ & $\asunc{2.40}{0.04}{0.05}$ & $\asunc{2.8}{0.1}{0.1}$ \\
    $C_{58}$& SF & $\asunc{2.4}{0.1}{0.1}$ & $\asunc{1.94}{0.04}{0.02}$ & $\asunc{2.1}{0.1}{0.1}$ \\
    & difference & $\asunc{0.3}{0.2}{0.2}$ & $\asunc{0.47}{0.05}{0.05}$ & $\asunc{0.7}{0.2}{0.2}$ \\
\hline
    & Q & $\asunc{0.76}{0.03}{0.04}$ & $\asunc{0.66}{0.01}{0.01}$ & $\asunc{0.73}{0.05}{0.07}$ \\
    $q$& SF & $\asunc{0.72}{0.04}{0.02}$ & $\asunc{0.57}{0.01}{0.01}$ & $\asunc{0.58}{0.05}{0.03}$ \\
    & ratio & $\asunc{1.1}{0.1}{0.1}$ & $\asunc{1.15}{0.03}{0.03}$ & $\asunc{1.3}{0.1}{0.2}$ \\
\hline
	\end{tabular}
\end{table}

\subsubsection{Radii}
In all simulations half-light radii of star-forming and quiescent galaxies are very similar, with a difference of the average size of at most $\SI{0.2}{dex}$. The median size of $\lmass > 11$ quiescent and star-forming galaxies in TNG300 is $\asuncunit{1.79}{0.07}{0.05}{kpc}$ and $\asuncunit{2.20}{0.04}{0.06}{kpc}$, respectively, with no significant dependence on stellar mass. Quiescent galaxies in \mthree\ have sizes of $\asunc{3.16}{0.39}{0.23}$, larger than those of star-forming galaxies with $\asuncunit{1.87}{0.18}{0.18}{kpc}$, though average sizes of star-forming and quiescent galaxies are consistent when estimated in 2 times the half-light radius apertures. The increase of average sizes of star-forming galaxies with stellar mass visible in \mthree\ in Figure~\ref{fig:morphology} is actually only seen for \SI{30}{kpc} aperture based sizes. If classifying galaxies based on UVJ restframe colors (see Section~\ref{sec:uvj}) average sizes of $\lmass>11$ quiescent and star-forming galaxy populations change by at most $\SI{0.05}{dex}$.

\cite{VanDerWel2014} study morphologies of star-forming and quiescent galaxies at $0<z<3$ based on data from the 3D-HST \citep{Brammer_2012_3DHST} and CANDELS \citep{Grogin2011_CandelsDesign, Koekemoer2011_CandelsData} surveys.
In their analysis, massive ($11.0<\lmass<11.5$) quiescent galaxies at $2.5<z<3.0$ are smaller than star-forming galaxies, with average sizes of $\asuncunit{2.5}{0.4}{0.5}{kpc}$ and $\asuncunit{3.55}{0.2}{0.2}{kpc}$, respectively (although with a large scatter, see Figure~\ref{fig:morphology}). The best-fit mass-size relations of quiescent and star-forming sources cross at $\lmass\approx 11.6$ because of their different slopes of $\dlogRdM = 0.79\pm0.07$ and $0.18\pm0.02$, respectively.
In \mthree\ quiescent galaxies have sizes consistent with observed ones from \cite{VanDerWel2014}.
However, at face value quiescent galaxy sizes in \mthree\ are similar in size to, or larger than, star-forming galaxies (depending on aperture choice). Nonetheless, because of the large statistical uncertainties, they are still consistent with having the same ratios as star-forming and quiescent galaxy sizes estimated by \cite{VanDerWel2014}.
In TNG300 quiescent galaxies have on average significantly smaller sizes than star-forming galaxies with a ratio of $0.7-0.8$ (depending on the considered aperture), in good quantitative agreement with measurements from \cite{VanDerWel2014}. The larger sample size in TNG300 also allows to investigate the stellar mass vs. size relation which is consistent with being flat in the probed $\lmass>11$ range for both star-forming and quiescent populations, at odds with the relation derived by \citet[][]{VanDerWel2014}.

\cite{Genel_2018_TNGSizeEvolution} investigate the mass-size relation of star-forming and quiescent galaxies with $\lmass>9$ in TNG100 up to $z=3$. At masses higher than $\lmass\approx 10.5$ and up to $z\approx2$, they find for both populations that the mass-size relation increases with a constant slope, in good quantitative agreement with determinations of the mass-size relation from \cite{Shen_2003}, \cite{Bernardi_2014_size_luminosity_relation_SDSS} and \cite{VanDerWel2014}. However, at $z=3$ \cite{Genel_2018_TNGSizeEvolution} find flat mass-size relations for both populations up to $\lmass=11$, in agreement with our results for TNG300 at higher masses. On the other hand, in a study in TNG simulations with higher resolution but smaller volume (TNG50), \citet[][]{Varma_2022_TNG50_sizes_scaling_relations} find sizes in good agreement with observed sizes from \cite{VanDerWel2014} up to $z=2.4$ and up to $\lmass\approx 11.5$. This potential inconsistency might come from different procedures adopted to estimate galaxy sizes (between the \cite{Genel_2018_TNGSizeEvolution} and \cite{Varma_2022_TNG50_sizes_scaling_relations} studies, and also with respect to this work) as well as possibly from resolution effects.

\cite{Remus_2017_morphology_feedback} investigated the mass-size relation for morphological early-type galaxies in Magneticum up to $z=2$. At all redshifts they find good agreement with \cite{VanDerWel2014} as well, down to stellar masses of $\lmass=10.7$, clearly showing that the observed evolution in the mass-size relation of early-type galaxies in Magneticum is reproduced successfully as well. However, from \cite{Teklu_2015_magneticum_angular_momentum_galaxy_dynamics} it is known that some of these early-type galaxies have still large gas fractions and thus are likely not quiescent.

\subsubsection{Specific angular momentum}
The average light-weighted projected angular momenta of $\lmass>11$ quiescent galaxies in \mthree\ and TNG300 are $\asunc{170}{40}{50}$ and $\asunc{230}{20}{20}$, respectively. In all simulations the average angular momentum is larger for star-forming than for quiescent galaxies, but the difference is only significant in TNG300 with a ratio of $0.55\pm 0.06$.

In Figure~\ref{fig:morphology} we compare the angular momenta from the simulations with results from \cite{Fall_2018_AngularMomentum} for disks and bulges. To account for the evolution of the specific angular momentum with redshift we extrapolate the \cite{Fall_2018_AngularMomentum} relations to $z\approx2.7$ based on results from \cite{Swinbank_2017_angular_momentum_evolution} and \cite{Lagos_2017_angular_momentum_evolution_EAGLE}. \cite{Swinbank_2017_angular_momentum_evolution} analyse observed star-forming galaxies at $0.3\lesssim z\lesssim 1.7$ and find that their angular momentum evolves with $(z+1)^{-1}$ ($\approx\SI{0.6}{dex}$ decrease between $z=0$ to $z=2.7$). To date observational statistical measurements of angular momenta of quiescent galaxies do not reach $z\approx 2$ and their redshift evolution is less constrained. However, by analysing galaxies in the \eagle\ simulations, \cite{Lagos_2017_angular_momentum_evolution_EAGLE} find evidence for a weaker evolution of the angular momentum of quiescent galaxies of only $\SI{0.2}{dex}$ between $z=0$ and 3 while star-forming galaxies show an evolution of $0.3-\SI{0.4}{dex}$ in the same redshift range. For an isolated spherical halo \cite{Obreschkow_2015_angular_momentum_scaling} find that the specific angular momentum evolves with $(1+z)^{-1/2}$ (increase by $\approx \SI{0.3}{dex}$ between $z=2.7$ and $z=0$) due to cosmic expansion. \cite{Teklu_2015_magneticum_angular_momentum_galaxy_dynamics} showed that the evolution of the specific angular momentum in Magneticum is consistent with the relation from \cite{Obreschkow_2015_angular_momentum_scaling}.
To account for the uncertainties in the expected redshift evolution of the angular momentum we therefore show the low redshift results from \cite{Fall_2018_AngularMomentum} scaled by $0.2$ and $\SI{0.6}{dex}$.
In TNG300 the average angular momentum of star-forming galaxies are in very good agreement with findings from \cite{Fall_2018_AngularMomentum} for disk galaxies scaled to $z\approx 2.7$ according to \cite{Swinbank_2017_angular_momentum_evolution}, while in \mthree\ the average angular momentum at $\lmass=11.1$ is lower by $\SI{0.2}{dex}$ and does not increase with stellar mass.
Angular momenta of quiescent galaxies with $\lmass\approx 11.1$ in TNG300 and \mthree\ are $\approx\SI{0.3}{dex}$ and $\approx\SI{0.1}{dex}$, respectively, higher than expected from the extrapolation to higher redshift from \cite{Lagos_2017_angular_momentum_evolution_EAGLE}. Although the angular momentum is expected to increase with stellar mass \citep[e.g.,][]{Fall_2018_AngularMomentum}, there is no significant trend in all simulations (albeit statistics is poor for M3 and TNG100).

\subsubsection{\bvalue}
\label{bvalue}
By analysing the specific angular momentum of galaxies in Magneticum up to $z\approx 2$ \cite{Teklu_2015_magneticum_angular_momentum_galaxy_dynamics, Teklu_2016_angular_momentum_inproceeding} found that disk and spheroidal galaxies populate different regions in the $\Mstar-j$ plane which therefore is an excellent tracer for morphology. Assuming the relation $j\propto\Mstar^{2/3}$ \citep[e.g.,][]{Romanowsky_2012_AngularMomentum} a single parameter, the \bvalue\ \citep{Teklu_2015_magneticum_angular_momentum_galaxy_dynamics}, can be used to separate between different morphological types.

In Figure~\ref{fig:bvalue} we show the estimated $b$-values of massive ($\lmass>11$) galaxies at $z\approx 2.7$ for all considered simulations, as a function of \ac{ssfr}. The solid red and blue lines define two regions in the diagram expected to be exclusively populated by early- and late-type galaxies (bottom and top part of the plot, according to \cite{Teklu_2015_magneticum_angular_momentum_galaxy_dynamics} based on the \cite{Obreschkow_2015_angular_momentum_scaling} scaling to $z=2.7$). Shaded areas around the blue and red solid lines show the impact of adopting a different scaling of angular momentum with redshift, from \cite{Lagos_2017_angular_momentum_evolution_EAGLE} and \cite{Swinbank_2017_angular_momentum_evolution}.

For M3 and TNG100 the average \bvalue\ is consistent with being flat as a function of sSFR, but statistics is poor.
For TNG300 where statistics is sufficient, Figure~\ref{fig:bvalue} shows a clear trend of \bvalue\ with \ac{ssfr}, spanning the range between the expectations for bulge- and disk-dominated galaxies.
However, Figure~\ref{fig:bvalue} also clearly shows that in all simulations the low-\bvalue\ region (below the solid red line), which is expected to be populated by morphologically early-type galaxies, actually contains sources over a broad range of sSFR, with a significant fraction of actively star-forming sources (only $\approx\SI{20}{percent}$ of these low-\bvalue\ galaxies are classified as quiescent in \mthree\ and
$\approx40-\SI{70}{percent}$ in TNG100 and TNG300.

\subsubsection{Axis ratios and concentration}
\label{sec:Concentration}
In all simulations we find that average projected axis ratios and concentrations of quiescent galaxies are larger than for star-forming galaxies (see lowest panels of Figure~\ref{fig:morphology}), however, due to the small sample sizes in \mthree\ and TNG100 these differences are only significant in TNG300.
Average axis ratios of quiescent galaxies in \mthree\ and TNG300 are $\asunc{0.76}{0.03}{0.04}$ and $\asunc{0.66}{0.01}{0.01}$, which is higher than for star-forming galaxies by a factor of $\asunc{1.08}{0.06}{0.07}$ and $\asunc{1.15}{0.03}{0.03}$. In \cite{Lustig_2021} we found for our observed sample of 10 quiescent galaxies an average axis ratio of $\asunc{0.73}{0.12}{0.06}$. The uncertainties on the average axis ratios of observed galaxies at high redshift are relatively large because of the small sample sizes and the additional dependence of the axis ratio on the inclination angle. We could therefore not see a significant difference between axis ratios of observed quiescent and star-forming galaxies.

The average concentration of the profiles of quiescent galaxies is larger than that of star-forming galaxies by $0.28\pm0.19$ and $0.47\pm0.05$ in \mthree\ and TNG300, respectively (see second lowest panels of Figure~\ref{fig:morphology}).
Although the differences in average axis ratio and concentration between quiescent and star-forming galaxies in the simulations are relatively small they suggest that the evolution of quiescent galaxies towards a spheroidal, bulge dominated structure is already in progress at this redshift. 
For comparison we calculate concentrations for galaxies from \cite{VanDerWel2014} with the same definition relying on results of their morphological analysis. We find a difference of the concentration of star-forming and quiescent galaxies of $\approx 0.8$, larger than found in the simulations, with no significant redshift trend.

\section{Summary and conclusions}
\label{sec:conclusion}

We have analysed massive ($\lmass > 11$) galaxies at $z\approx 2.7$ in the Magneticum (\mthree) and IllustrisTNG (TNG100 and TNG300) hydrodynamical simulations in the GAEA semi-analytic model, and compared with observational results on stellar population and structural properties at similar redshift. In our analyses we considered different prescriptions for dust attenuation for quiescent and star-forming galaxy populations and different criteria for defining quiescence as well as a range of apertures for galaxies in hydrodynamical simulations, to estimate the impact of such assumptions on the comparison of simulations and observations carried out in this work.

We investigate the main sequence of star-forming galaxies in the studied simulations and find that in \mthree\ and GAEA the \ac{sfr} increases constantly with stellar mass with a slope close to unity (corresponding to $\lssfr\approx-9.3$ and $-8.8$ in \mthree\ and GAEA, respectively) over the full mass range. In both TNG simulations at $\lmass\gtrsim 11.1$ a large fraction of galaxies is quenched and we can only identify a \ac{ms} in TNG300 at lower masses.
Observed main sequences from \cite{Sargent_2014_main_sequence} and \cite{Schreiber_2015} have a slope in agreement with our determination for GAEA, \mthree\ and TNG300 (although there is a strong bending at $\lmass>11.1$ in TNG300) but are offset to higher \ac{sfr} by $\approx\SI{0.6}{dex}$ compared to \mthree\ and TNG300. We stress that the lack of a clear bimodality in the hydrodynamical simulations and the high fraction of galaxies with suppressed star formation especially in TNG simulations at high masses may bias our determination of the \ac{ms}.
Following inspection of the star formation rate vs. stellar mass diagram in the different simulations (see Section~\ref{sec:sampleselection}) we thus adopt the \cite{Franx_2008} criterion to identify quiescent galaxies in all studied simulations. The criterion defines a \ac{ssfr} threshold for quiescence and does therefore not depend on the actual determination of the \ac{ms} in the individual simulations investigated here. However, it turns out to be roughly equivalent to defining quiescent galaxies as those with a \ac{sfr} $\SI{1}{dex}$ below the \ac{ms} (see Section~\ref{sec:quiescent_selection}).

Based on this selection, as shown in Figure~\ref{fig:mainsequence}, for \mthree\ and GAEA the quiescent fractions ($\SI{19}{percent}$ and $\SI{27}{percent, respectively,}$ at $\lmass>11)$ are anyway largely consistent with most observations \citep[except][see Section~\ref{sec:passivefractions}]{Davidzon_2017_SMF_COSMOS2015}.
In TNG simulations, quiescent fractions are typically larger than observed: while still close to observed values in TNG100 (albeit with poor statistics), they can reach twice as high as observed in TNG300.
If quiescent galaxies are identified by UVJ colors instead, quiescent fractions at $\lmass>11$ in \mthree\ increase to $\SI{30}{percent}$ while quiescent fractions in TNG simulations differ by $<5$ percentage points from the ones estimated with the \ac{ssfr} criterion.

Average ages in terms of $t_{50}$ for the quiescent galaxy population in all simulations are $\gtrsim \SI{0.8}{Gyr}$. Although the distributions of ages still overlap, this is significantly older than the average $t_{50}$ of $\SI{0.5}{Gyr}$ of observed galaxies at similar redshift. Because of the few spectroscopically confirmed quiescent galaxies, this comparison is limited to the 10 galaxies from \cite{DEugenio_2020_paper}.
In \cite{Lustig_2021} and \cite{DEugenio_2020_paper} we analysed the massive parent sample of quiescent galaxies at $2.5<z<3.0$ in the \cite{Muzzin_2013} and \cite{LaigleCOSMOS2016} catalogs and found a potential mild bias of the studied spectroscopically confirmed sample towards younger stellar ages, due to the selection of quiescent candidates for spectroscopic follow-up.
However, this selection effect alone cannot explain the discrepancy between the ages in the simulations vs. observations (see Section~\ref{sec:stellarages}). By analysing the \acp{sfh} of quiescent galaxies in the simulations we find that a significant fraction of their stellar mass is already formed at early times during a relatively slow increase of the \ac{sfr}, and a higher sensitivity to the more recent \ac{sfh} in spectro-photometric modeling may cause an underestimation of observed ages. Indeed, photometric modeling of synthetic photometry for simulated galaxies with the same approach as in \cite{DEugenio_2020_paper} suggests that derived age estimates may be underestimated, although this cannot fully account for the discrepancy. We also note that spectroscopic studies of (38, overall) quiescent galaxies at $3\lesssim z < 4$ from \citet[][]{Gobat_2012, Schreiber_2018_NIR_spec_3z4, Saracco_2020_rapid_buildup_massive_ET_z3.4, Valentino_2020_Q_1.5years_after_binbang_and_progenitors, Forrest_2020_one_massive_galaxy_z3.5, Forrest_2020_survey_rapid_SF_and_quenching_early_universe, Carnall_2022_JWST_Q_galaxies} have revealed similarly young ages as in \cite{DEugenio_2020_paper}; however, if assuming pure passive evolution once these galaxies have quenched, the ages of their descendants at $z\approx 2.7$ would be comparable to those of simulated galaxies investigated here. However, it remains unclear how representative these potential older descendants are for the overall population of $z\approx 2.7$ quiescent galaxies.
For these reasons, a quantitative estimate of the discrepancy between stellar ages in simulated vs. observed quiescent galaxies at $z\approx 3$
is affected by potential observational biases and inconsistencies in the analysis of simulated vs. observed sources.

We investigate the restframe UVJ color plane of simulated galaxies adopting recipes for dust attenuation of both quiescent and star-forming sources, and photometric uncertainties typical of deep field surveys used for studies at this redshift. We find that, according to our analysis of simulated galaxies, UVJ quiescent samples would be strongly contaminated by dusty star-forming sources in the UVJ region typically populated by the oldest quiescent galaxies, reducing the overall purity of a UVJ selected sample to $\approx40-\SI{50}{percent}$ according to our modeling \citep[in qualitative agreement with our analysis of observations in][]{Lustig_2021}.
In agreement with results from previous studies \citep[e.g.,][]{Merlin_2018, Schreiber_2018_NIR_spec_3z4, Santini_2021_emergence_passive_galaxies_early_univ} we find that the UVJ selection with the routinely adopted criteria \citep[e.g.,][]{Williams_2009} leads to incomplete quiescent galaxy samples at this redshift, since a non negligible fraction of recently quenched quiescent galaxies with significantly suppressed \ac{sfr} (in this case e.g. $\lssfr<-10$) has not yet entered the UVJ quiescent region due to the $U-V > 1.3$ constraint.

We finally investigate structural properties of quiescent galaxies in the considered hydrodynamical simulations, although because of the relatively small sample size in TNG100 and \mthree\ we can detect structural differences between quiescent and star-forming populations at a significant level only in TNG300.
We find that simulated quiescent galaxies are on average more centrally concentrated and have higher stellar axis ratios than star-forming galaxies, indicating that morphological differences are already emerging at $z\approx 3$.
In all simulations the ratio of quiescent and star-forming galaxy sizes in the observed H-band at $\lmass>11$ is formally in agreement with observations at similar redshift \citep[][but note poor statistics affecting \mthree\ and TNG100 results]{VanDerWel2014}. With the larger sample size in TNG300 we also investigate the dependence of sizes on stellar mass and find the size-mass relation consistent with being flat for both populations in the probed $\lmass>11$ range.

Due to the complications hampering a proper, direct and fair comparison of structural properties in observations vs. simulations in our analysis (see full discussion in Section~\ref{sec:morphology}), it is not trivial to establish in absolute terms whether the correlation between structural and stellar population properties already seen at this redshift in several studies (see \citealt[][]{Bell_2012, Lang_2014, Tacchella_2015, Mowla_2019_b, Esdaile_2020, Lustig_2021}, but see also
\citealt[][]{VanDokkum_2008, McGrath_2008, Bundy_2010, VanDerWel_2011, Chang_2013, McLure_2013, Hsu_2014, Bezanson_2018} for a different picture) is actually quantitatively reflected in simulated galaxies.
In fact, based on the analysis of \bvalue s, that have been shown to correlate with galaxy morphology \citep[][]{Teklu_2015_magneticum_angular_momentum_galaxy_dynamics, Teklu_2016_angular_momentum_inproceeding}, early-type morphology and quiescence do not seem to be necessarily tightly related in the studied simulations with a significant fraction of morphologically early-type sources being still actively star forming (see Section~\ref{sec:morphology_discussion}). A more specific investigation of the early paths of structural evolution in connection with quenching in the studied simulations will be discussed in a future work.

\section*{Acknowledgements}

The IllustrisTNG simulations were undertaken with compute time awarded by the Gauss Centre for Supercomputing (GCS) under GCS Large-Scale Projects GCS-ILLU and GCS-DWAR on the GCS share of the supercomputer Hazel Hen at the High Performance Computing Center Stuttgart (HLRS), as well as on the machines of the Max Planck Computing and Data Facility (MPCDF) in Garching, Germany.
KD acknowledges support by the COMPLEX project from the European Research Council (ERC) under the European Union’s Horizon 2020 research and innovation program grant agreement ERC-2019-AdG 882679 as well as support by the Deutsche Forschungsgemeinschaft (DFG, German Research Foundation) under Germany’s Excellence Strategy - EXC-2094 - 390783311. The Magneticum Simulations
were carried out at the Leibniz Supercomputer Center (LRZ) under the project pr86re.
This work was supported by JSPS KAKENHI Grant Number JP21K03622.

\section*{Data availability}
The IllustrisTNG simulations are publicly available and accessible at www.tng-project.org/data \citep[][]{Nelson_2019}.
The Magneticum simulations and data directly related to this publication and its figures are available upon request from the corresponding author.



\bibliographystyle{mnras}
\bibliography{biblio} 




\appendix
\section{Impact of sSFR threshold for quiescence on quiescent fractions}
\begin{figure*}
	\includegraphics[width=\textwidth]{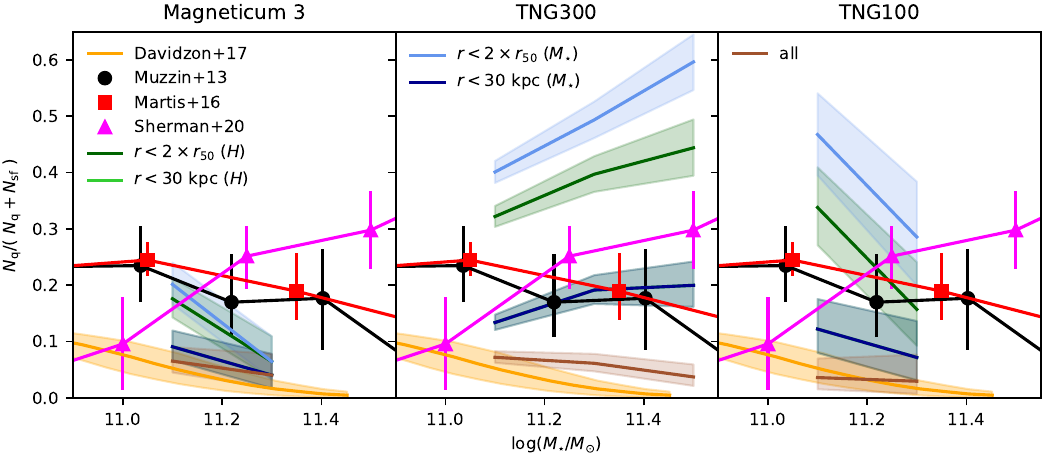}
	\caption{Quiescent fractions in the simulations for all apertures considered in this work (see Section~\ref{sec:sampleselection}) adopting a threshold of $\lssfr = -11$ to define quiescence.
	We compare with observational results from \protect\cite{Muzzin_2013_SMF}, \protect\cite{Martis_2016_passive_fractions}, \protect\cite{Davidzon_2017_SMF_COSMOS2015} and \protect\cite{Sherman_2020_passive_fractions} as indicated in the legend.}
	\label{fig:appendix_passive_fractions}
\end{figure*}

In Section~\ref{sec:sampleselection} we discuss different \ac{ssfr} thresholds for defining quiescence in simulations. To take into account the evolution of the normalisation of the \ac{ms} with redshift \citep[e.g.,][]{Speagle_2014_main_sequence, Johnston_2015_SFR_stellar_mass_since_z3, Schreiber_2015, Tomczak_2016_main_sequence_since_z4} we adopt for the analyses in this work a redshift dependent \ac{ssfr} threshold ($\mathrm{sSFR}\approx \SI{1E-10}{yr^{-1}}$ at $z=2.7$) to define quiescence. In Figure~\ref{fig:appendix_passive_fractions} we show quiescent fractions if a \ac{ssfr} threshold of $\SI{1E-11}{yr^{-1}}$, often used for classification at low redshift, is used instead and compare with observational results from \cite{Muzzin_2013_SMF}, \cite{Martis_2016_passive_fractions}, \cite{Davidzon_2017_SMF_COSMOS2015} and \cite{Sherman_2020_passive_fractions}.
At $\lmass>11$ we find overall quiescent fractions between $5$ and $\SI{15}{percent}$ for \mthree, lower than found by \cite{Muzzin_2013_SMF}, \cite{Martis_2016_passive_fractions} and \cite{Sherman_2020_passive_fractions} but still higher than in \cite{Davidzon_2017_SMF_COSMOS2015}.
Quiescent fractions in both TNG simulations in the same mass range show a very strong aperture dependence with overall quiescent fractions between $7$ and $\SI{45}{percent}$ in TNG300 and $2$ and $\SI{43}{percent}$ in TNG100. The highest quiescent fractions are measured for an aperture of 2 times the half-mass radius, the lowest if all bound particles are considered.

\begin{table*}
	\centering
	\caption{Properties of the simulations used for this work.}
	\label{tab:simulations_general}
	\begin{tabular}{lcccc}
		\hline
		 & Magneticum 3 & TNG300 & TNG100 & GAEA\\
		\hline
		$m_{\mathrm{b}}\,[10^6\,\Msun]$ &7.3&11.0&1.4&--\\
		$m_{\mathrm{DM}}\,[10^6\,\Msun]$ &36.0&59.0&7.5&--\\
		$\epsilon_{\star}$ [comoving kpc/h]  & 0.7 & 2.0 & 1.0&--\\
		Box size [comoving Mpc] & 182 & 303 & 111 & 685\\ \hline
		\multicolumn{5}{c}{Cosmology} \\
		$\Omega_{\mathrm{M}}$ 				    & 0.272 &\multicolumn{2}{c}{0.309}&0.25\\
		$\Omega_{\Lambda}$    & 0.728 &\multicolumn{2}{c}{0.691}&0.75\\
		$\omegabaryon$& 0.046 &\multicolumn{2}{c}{0.049}&0.045\\
		h                                    & 0.704 &\multicolumn{2}{c}{0.677}&0.73\\
		$\sigma_8$                   & 0.809 &\multicolumn{2}{c}{0.816}&0.9\\
		\hline
		\multicolumn{5}{c}{The sample} \\
		$\zsnap$&2.79&\multicolumn{2}{c}{2.73}&2.83\\
		$\epsilon_{\star,\zsnap}\,$[physical kpc] & 0.26 & 0.8 & 0.4&--\\
		\multicolumn{5}{c}{Number of $\lmass>11$ galaxies ($\SI{30}{kpc}$ aperture for hydrodynamical simulations)} \\
		All & 166 & 993 & 73 & 9339\\
		Star-forming & 136 & 549 & 47 & 2561\\
		Quiescent & 30 & 444 & 26 & 6778\\
		\hline
	\end{tabular}
\end{table*}

\section{Impact of a higher sSFR threshold for defining quiescence on estimated purity and completeness of UVJ selected samples}
\begin{figure*}
	\includegraphics[width=\textwidth]{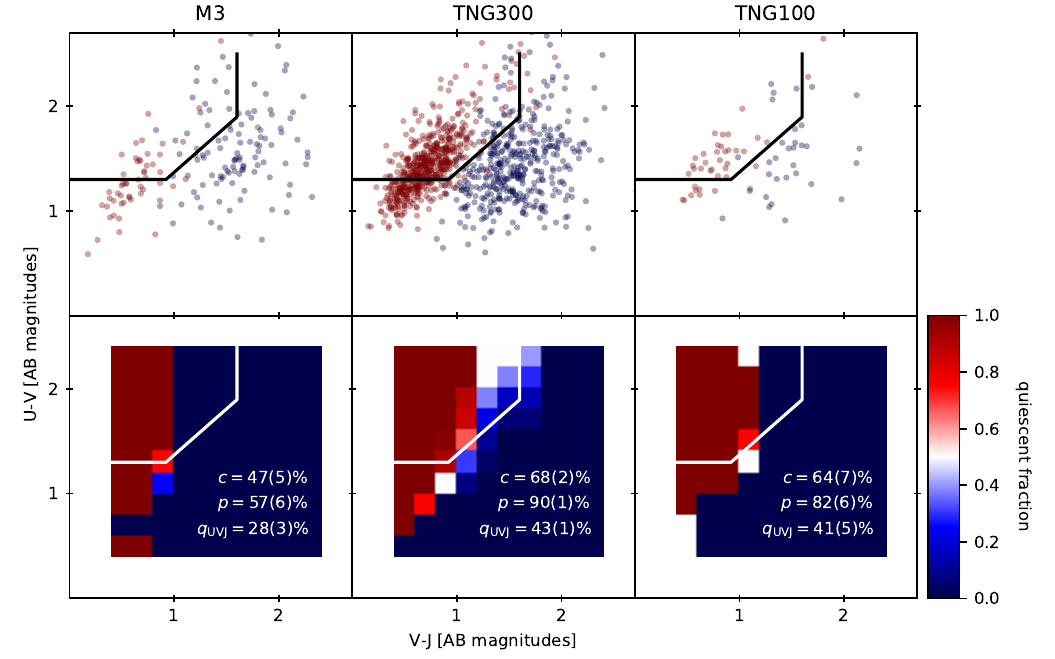}
	 \caption{Restframe $U-V$ vs. $V-J$ colors for simulated galaxies.
	 In the upper panel we show a random single realisation of the colors after applying dust attenuation. In the middle panel we show the average fraction of quiescent galaxies from 1000 realisations as a function of the position in the UVJ diagram as indicated by the colorbar if a threshold of $\mathrm{sSFR}=0.75\times H(z)$ is adopted.
	 The numbers show completeness ($c$) and purity ($p$) of the overall UVJ-quiescent sample and the quiescent fraction according to UVJ selection ($q_{\mathrm{UVJ}}$) together with the scatter across the different realizations.}
	 \label{fig:appendix_uvj_relaxed_ssfr_cut}
\end{figure*}

In Section~\ref{sec:uvj} we analyse purity, completeness and UVJ-derived quiescent fractions for a UVJ selected sample of quiescent galaxies. In Figure~\ref{fig:appendix_uvj_relaxed_ssfr_cut} we show UVJ diagrams for simulated massive galaxy samples where we adopted a relaxed \ac{ssfr} threshold for defining quiescence ($\mathrm{sSFR} = 0.75\times H(z)$, see Section~\ref{sec:stellarages}). The different threshold for quiescence affects the implementation of dust attenuation for the individual galaxies, and thus the galaxy distribution in the UVJ diagram. Ultimately, the adopted definition of quiescence affects by construction the purity and completeness of the UVJ-selected quiescent samples as determined in Section~\ref{sec:uvj}. We find overall UVJ-derived quiescent fractions of $\SI{28}{percent}$ in \mthree\ and $\approx\SI{40}{percent}$ in the TNG simulations, consistent with UVJ-derived quiescent fractions obtained adopting the \cite{Franx_2008} criterion (as in Section~\ref{sec:uvj}). The estimated completeness of UVJ-selected quiescent samples in the considered mass range decreases by $10-20$ percentage points with respect to our results in Section~\ref{sec:uvj} because a significant fraction of galaxies classified as quiescent with the relaxed threshold are very young and have not yet entered the UVJ quiescent region. The purity increases by $10-15$ percentage points because a larger fraction of galaxies in the UVJ quiescent region is defined as quiescent with the relaxed \ac{ssfr} threshold.

\bsp	
\label{lastpage}
\end{document}